 \documentclass[keywords,a4paper,11pt]{article}
\usepackage{amsmath}
\usepackage{amssymb}
\usepackage{amsfonts}
\usepackage{latexsym}
\usepackage{epsfig}
\usepackage{float}
\usepackage{a4wide}
\usepackage{hhline}
\usepackage{rotate}
\makeatletter
\renewcommand\@biblabel[1]{#1.}
\makeatother
\begin{document}

\title{\bf{Use of specific Green's functions for solving direct problems involving a heterogeneous rigid frame porous medium slab solicited by acoustic waves}}
\author{J.-P. Groby $^{1}$\thanks{Correspondence to: J.-P. Groby, Akoestieke en Thermische Fysica, KULeuven, Celestijnenlaan 200D , 3001 Heverlee, Belgium} \thanks{E-mail: jeanphilippe.groby@fys.kuleuven.be}, L. De Ryck $^{1}$, P. Leclaire $^{1}$, A. Wirgin $^{2}$,
\\
W. Lauriks $^{1}$, R.P. Gilbert $^{3}$ and Y.S. Xu $^{4}$
\\[12pt]
$^{1}$ \textit{Akoestieke en Thermische Fysica, KULeuven},\\
\textit{Celestijnenlaan 200D , 3001 Heverlee, Belgium}\\[8pt]
$^{2}$ \textit{Laboratoire de M\'ecanique et d'Acoustique, UPR7051 du CNRS},\\ 
\textit{31 chemin Joseph Aiguier, 13402 Marseille cedex 20, France}\\[8pt]
$^{3}$ \textit{Mathematical Sciences, University of Delaware, Newark,
Delaware 19716, USA}\\[8pt]
$^{4}$ \textit{Mathematics Department, University of Louisville,
Louisville, KY 40292, USA}}
\date{}
%\keywords{Integral method, Born approximation, specific Green's function, macroscopically inhomogeneous porous medium}
\maketitle
%Ajout
\begin{abstract}
A domain integral method employing a specific Green's function
(i.e., incorporating some features of the global problem of wave
propagation in an inhomogeneous medium) is developed for solving
direct and inverse scattering problems relative to slab-like
macroscopically inhomogeneous porous obstacles. It is shown how to
numerically solve such problems, involving both spatially-varying
density and compressibility, by means of an iterative scheme
initialized with a  Born approximation. A numerical solution is
obtained for a canonical problem involving a two-layer slab.
\end{abstract}
%%%%%%%%%%%%%%%%%%%%%%%%%%%%%%%%%%%%%%%%%%%%%%%%%%%%%%%%%%%%%%%%%%%%%
\section{Introduction}

This work was initially motivated by two problems: i) the  design
problem connected with the determination of the optimal profile of
a continuous and/or discontinuous spatial distribution of the
material/geometric properties of porous materials for the
absorption of sound  \cite{boro98} and ii) the retrieval of the
spatially-varying mechanical and geometrical parameters of bone
for the diagnosis of diseases such as osteoporosis \cite{bugi03}.

Such inverse problems \cite{hah02} can be decomposed into
 two sub-problems: i) the determination of the constitutive and
 conservation relations linking the
various spatially-variable mechanical parameters of the porous
medium to its response to an acoustic solicitation, and ii) the
resolution of the wave equation in an inhomogeneous porous medium
(for instance, within the Biot, or rigid frame approximations).
Here we focus on the second point.

In \cite{Deryck}, it is shown that the wave equation describing
the propagation in a macroscopically-inhomogeneous porous medium
in the rigid frame approximation can formally take the form of the
usual acoustic wave equation in a macroscopically-inhomogeneous
fluid (in which the microscopic features of the  porous medium are
homogenized) with spatial (and frequency) dependent
compressibility $\kappa_{e}(\mathbf{x},\omega)$ and density
$\rho_{e}(\mathbf{x},\omega)$.

The present work deals with a method of resolution of direct
problems involving acoustic wave propagation in a
macroscopically-inhomogeneous fluid medium, whose density and
compressibility are both space dependent, this being a
prerequisite to the resolution of related inverse problems.

This topic is also of great interest in quantum physics (inverse
potential scattering \cite{aktoklaus02,newt02,sab00,sab02}), ocean
acoustics \cite{breklys91,brek60,lamoli92,bugi98,munwar95,bugi04}
(detection of inhomogeneities, sediment exploration, influence of
seawater and seafloor composition and heterogenity on the
long-range propagation of acoustic waves in the sea, ...),
seismology \cite{akirich80,scales02,snie99} (determination of the
internal structure and composition of the Earth via seismic
waves,...), geophysics \cite{snie99,jondem96,zhumcm91,wap96}
(characterization of soil, detection of geological features such
as hydrocarbon reservoirs, ...), optics and electromagnetism
\cite{yablo94,smipad00} (design and characterization of materials
having specified response to waves, detection of flaws,...).

The wave equation in an inhomogeneous medium can be solved in a
variety of manners: via the wave splitting method
\cite{PIER,Deryck,kriskarl02}, the transfer matrix method
\cite{knop64,Allard} (for piecewise constant media), integral
methods \cite{hah02,sab02,newt02}, or purely numerical (e.g.,
finite-element \cite{grts03} or finite-difference \cite{alter68})
methods. The methods dedicated to inverse problems are wave
splitting  and linearisation \cite{sab00,sab02} techniques
deriving from the integral formalism. The two most widely-known
approximations for the Fredholm equations of the second kind
involved in the integral formalism (at least when the density is
constant in the acoustic context) are the Born approximation
\cite{Morse,sab02} and the Rytov approximation \cite{ishimaru}. We
will focus on the Born approximation, despite the fact that
several authors \cite{keller,chernov} have shown that the Rytov
approximation is valid under a less restrictive set of conditions
than the Born approximation.

We  postulate, and show, that the accuracy of the Born
approximation can be increased by the use of the integral
formulation together with a specific Green's function
\cite{newt02,sab02,wirgin2006}. In most of the articles dealing
with the Born approximation and other linearisation methods, the
problems are often simplified by considering the density to be
constant. Herein, we consider both the compressibility and the
density to be spatially-variable. This induces supplementary
difficulties, because it can lead to meaningless integrals
(involving first and/or second space derivatives of the density),
especially when the variation of the density is not continuous,
and/or because it requires the evaluation of the first space
derivative of the pressure field. We will show how to deal with
these problems.

The usual first-order Born approximation is an outcome of the
integral formulation employing the free-space Green's function
(FSGF) and consists in approximating the pressure field in the
integrand (corresponding to the pressure field inside the
heterogeneity) by the field in the absence of the heterogeneity
(i.e. the incident field), this being equivalent to the
asssumption that the diffracted field is negligible compared to
the incident field. Although this method usually provides good
results for small contrasts between the mechanical parameters of
the inhomogeneity and those of the host medium, its accuracy
decreases in the case of dissipative media and larger contrasts.

Moreover, iterative schemes initialized with the zeroth-order Born
or Rytov approximations often diverge in practice. This difficulty
can be partially resolved by employing the modified or distorted
Born approximation \cite{sab02} which basically consist in acting
on one of the three terms involved in the integrand (the Green's
function, the contrast function and the field inside the
heterogeneity).

The method described herafter allows us to act on all  three
terms. The central idea of our method consists in reducing the
eigenvalues of the kernel of the integrand (thought to be the
cause of the difficulties with the usual iterative Born scheme) by
employing a Green's function--the so-called Specific Green's
function (SGF)-- of a  canonical problem which is close, in some
sense, to the original problem. The  specification of the initial
solution was already treated in \cite{wirgin2006} in connection
with the resolution of an inverse problem. The chosen problem
(also canonical) is that of the diffraction of an incident plane
wave,
 propagating  in the host medium, by a two-component slab (each
 component being a homogeneous layer)
considered as a single inhomogeneous slab. In the present instance, the
close canonical problem involves a slab filled with a
macroscopically-homogeneous fluid-saturated porous medium
surrounded by the same fluid (air) medium as the original
macroscopically-inhomogeneous fluid-saturated porous medium.

We will show, for this example: i) how to construct an appropriate
specific Green's function (SGF), ii) how to incorporate the latter
into the integral formulation, and iii) how the resulting integral
equation can be solved by an iterative scheme initialized by a
modified Born approximation.

The results are compared to those of the analytic solutions
(obtained by the transfer matrix method (TMM)) for the
two-component slab and found to be in good agreement with the
latter, both in transmission and reflection and for several angles
of incidence. The iterative scheme initialized with the modified
Born approximation converges rather rapidly, i.e., within $5$ to
$7$ iterations for our example. This demonstrates the efficiency
of our SGF interative scheme  for the resolution of the direct
problem relative to wave propagation in the presence of a
macroscopically-inhomogeneous  fluid-saturated porous slab.
%%%%%%%%%%%%%%%%%%%%%%%%%%%%%%%%%%%%%%%%%%%%%%%%%%%%%%%%%%%%%%%%%%%%%
\section{Use of the specific Green's functions in the domain integral
formulation to solve direct scattering problems}\label{section1}
%--------------------------------------------------------------------
\subsection{An example of a direct scattering
problem}\label{section11}
\begin{figure}[H]
\centering\psfig{figure=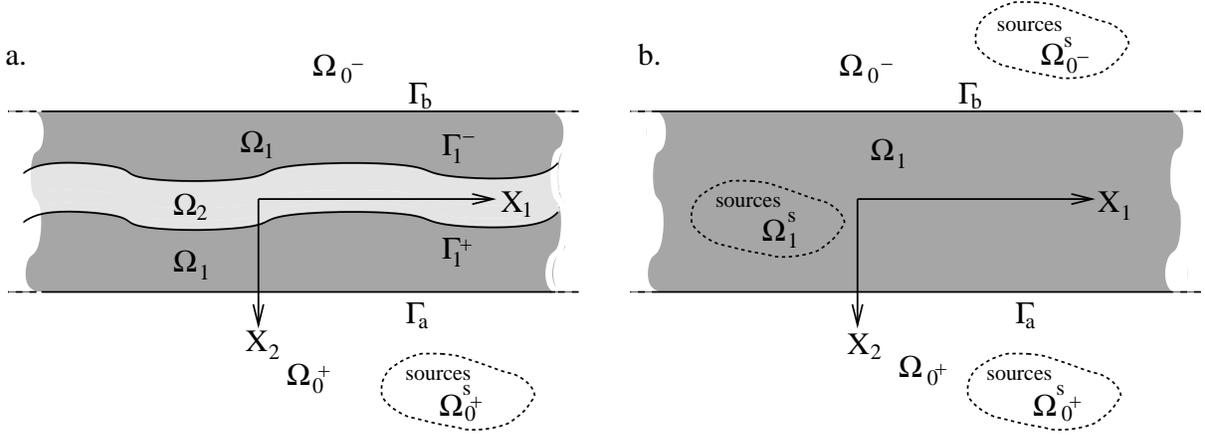,width=16.0cm}
\caption{Configuration of the direct problem of a.: a fluid heterogeneity within a fluid-like slab, b.: diffraction of a wave radiated by applied sources from a fluid slab of boundaries $\Gamma_a$ and $\Gamma_b$ immersed in a fluid.}
\label{section11figure1}
\end{figure}

The type of direct problem we deal with is illustrated in
figure \ref{section11figure1}a. This problem involves spatially-
dependent compressibility and density. As will be shown further
on, the spatial variability, and discontinuity, of both these
quantities, and in particular of the latter, can produce some
difficulties in the domain integral formulations (but not in the
TMM formulation). The wave equation for such problems is given in
appendix \ref{appendixe2}; to solve them in optimal manner, we
treat the auxiliary problem depicted in
figure \ref{section11figure1}b.
%%
%\begin{minipage}{8.0cm}
%%%%%%%%%%%%%%%%%%%%%%%%%%%%%%%%%%%%%%%%%%%%%%%%%%%%%%%%%%%%%%%%%%%%%
\subsection{Specific Green's function corresponding to the propagation of waves radiated
by interior and exterior line sources in the presenece of a
homogeneous fluid-like layer immersed in a homogeneous fluid
host medium}\label{section12}
%--------------------------------------------------------------------
\subsubsection{Features of the problem}\label{section121}
The sagittal plane (cross-section) view of the scattering
configuration is given in figure \ref{section11figure1}b. As we are
dealing with a Green's function, the supports of the sources
reduce to dots in the figure, i.e., the sources are line sources.
The homogeneous fluid-like layer is oriented horizontally (i.e.
the normal to both of its faces is along the $x_{2}$ axis); its
thickness is $l$, and the medium $M^1$ therein is homogeneous. The
geometry and composition of the layer are thus invariant with
respect to $x_3$. $\Omega_1$ designates the trace of the layer in
the $x_1-x_2$ cross-section plane. $\Gamma_a$ and $\Gamma_b$
designate the traces of the lower and upper faces respectively of
the layer in the $x_1-x_2$ cross-section plane. The unit vectors
normal to $\Gamma_a$ and $\Gamma_b$ are designated indistinctly by
$\boldsymbol{\nu}$. The $x_2$ coordinates of $\Gamma_a$ and
$\Gamma_b$ are designated by $a$ and $b$ respectively.

The layer is immersed in a (host) fluid $M^0$. The trace of the
host medium domain below (above) the layer in the $x_1-x_2$
cross-section plane is designated by $\Omega_{0^+}$
($\Omega_{0^-}$).

The (direct scattering) problem is to determine the response
$g^{0^{+}}$ within $\Omega_{0^+}$, $g^{0^{-}}$ within
$\Omega_{0^-}$ and  $g^{1}$ within $\Omega_{1}$ for line sources
that are located either within $\Omega_{0^+}$, $\Omega_{0^-}$ or
$\Omega_1$. This response constitutes the specific Green's
function we are looking for.

Let $\mathbf{y}$ designate the vector from $O$ to the location of
the line source. The Green's function in $\Omega_j$ is designated
by $g^j(\mathbf{x},\mathbf{y})$, which means the response at
$\mathbf {x}$ due to line sources located at $\mathbf{y}$.
%--------------------------------------------------------------------
\subsubsection{Governing equations}\label{section122}
Rather than to solve directly for $g^{j}(\mathbf {x},\mathbf{y},t)$, we prefer to deal with its Fourier transform $g^{j}(\mathbf {x},\mathbf{y},\omega)$ defined by:
\begin{equation}
g^{j}(\mathbf {x},\mathbf{y},t)=\int_{-\infty}^{\infty}g^{j}(\mathbf {x},\mathbf{y},\omega)e^{-\mbox{i} \omega t}d \omega\mbox {; }j=0^{+},1,0^-
\label{section122e1b}
\end{equation}
The mathematical translation of the boundary-value problem in the space-frequency domain is:
\begin{equation}
\left[\triangle +\left(k^j \right)^2
\right]g^j\left(\mathbf{x},\mathbf{y},\omega\right)=
-\delta(\mathbf{x}-\mathbf{y})\mbox{ : }\forall \mathbf{x}\in
\Omega_j\mbox{, }j=0^+,1,0^-\mbox{, } \mathbf{y}\in
\Omega_{0^+}\mbox{, }\Omega_{1}\mbox{ or }  \Omega_{0^-},
\label{section122e1}
\end{equation}
\begin{equation}
\begin{array}{ll}
\displaystyle \forall \mathbf{x}\in \Gamma_a &\displaystyle
\left\{
\begin{array}{l}
\displaystyle
g^{0^+}\left(\mathbf{x},\mathbf{y},\omega\right)-g^1\left(\mathbf{x},\mathbf{y},\omega\right)=0\\[8pt]
\displaystyle \frac{1}{\rho^0} \boldsymbol{\nu} (\mathbf{x})
\cdot\nabla
g^{0^+}\left(\mathbf{x},\mathbf{y},\omega\right)-\frac{1}{\rho^1}
\boldsymbol{\nu} (\mathbf{x}) \cdot\nabla
g^{1}\left(\mathbf{x},\mathbf{y},\omega\right)=0
\end{array}
\right.
\end{array}
\label{section122e2}
\end{equation}
\begin{equation}
\begin{array}{ll}
\displaystyle \forall \mathbf{x}\in \Gamma_b &\displaystyle
\left\{
\begin{array}{l}
\displaystyle
g^{1}\left(\mathbf{x},\mathbf{y},\omega\right)-g^{0^-}\left(\mathbf{x},\mathbf{y},\omega\right)=
0\\[8pt] \displaystyle \frac{1}{\rho^1} \boldsymbol{\nu}
(\mathbf{x}) \cdot\nabla
g^{1}\left(\mathbf{x},\mathbf{y},\omega\right)- \frac{1}{\rho^0}
\boldsymbol{\nu} (\mathbf{x}) \cdot\nabla
g^{0^-}\left(\mathbf{x},\mathbf{y},\omega\right)=0
\end{array}
\right.
\end{array}
\label{section112e3}
\end{equation}
\begin{equation}
g^j\left(\mathbf{x},\mathbf{y},\omega\right)-G^j\left(\mathbf{x},\mathbf{y},\omega\right)\sim
\mbox{ outgoing waves, }\mbox{ ; }\forall
\mathbf{x}\in\Omega_j\mbox{ ; }j=0^+,1,0^-\mbox{, }\|\mathbf{x}\|
\rightarrow \infty \label{section122e5}
\end{equation}
wherein $G^{j}$ is the free-space Green's function in the medium
$M^j$ given by
\begin{equation}
G^{j}(\mathbf{x},\mathbf{y},\omega)=\frac{i}{4}H_{0}^{(1)}(k^{j}(\|
\mathbf{x}-\mathbf{y}\|)=
\frac{\mbox{i}}{4\pi}\int_{-\infty}^{\infty}\exp\left(\mbox{i}k_1\left(x_1-y_1
\right)+
\mbox{i}k_2^j\left|x_2-y_2\right|\right)\frac{dk_1}{k_2^j}~,
\end{equation}
with $H_{0}^{(1)}$ the zeroth-order Hankel function of the first
kind, $k_2^{j}=\sqrt{\left(k^j\right)^2-(k_1)^2}$ such that
$\Re\left(k_2^{j} \right)\geq0$ and $\Im\left(k_2^{j}
\right)\geq0$, $j=0,1$ for $\omega\geq 0$.
%--------------------------------------------------------------------
\subsubsection{Field representations}\label{section123}
We shall henceforth: i) drop the $\omega$-dependence with the
understanding that it is implicit in all the field functions and
ii) employ the cartesian coordinates $(x_1,x_2)$ of $\mathbf{x}$
and $(y_1,y_2)$ of $\mathbf{y}$.

We use the separation of variables technique to obtain
\begin{equation}
g^{0^+}(\mathbf{x},\mathbf{y})=\mbox{H}_{\Omega_{0^+}}(\mathbf{y})G^{0^+}
(\mathbf{x},\mathbf{y})+\int_{-\infty}^{\infty}
B^{0^+}\exp\left(\mbox{i}k_1x_1+\mbox{i}k_2^{0}\left(x_2-a
\right)\right)\frac{dk_1}{k_2^0} \label{section123e1}
\end{equation}
\begin{equation}
g^{1}(\mathbf{x},\mathbf{y})=\mbox{H}_{\Omega_{1}}(\mathbf{y})G^{1}
(\mathbf{x},\mathbf{y})+\int_{-\infty}^{\infty} \left(
A^{1}\exp\left(-\mbox{i}k_2^{1}x_2\right)+
B^{1}\exp\left(\mbox{i}k_2^{1}x_2\right)\right)
\exp\left(\mbox{i}k_1x_1\right)\frac{dk_1}{k_2^1}
\label{section123e2}
\end{equation}
\begin{equation}
g^{0^-}(\mathbf{x},\mathbf{y})=\mbox{H}_{\Omega_{0^-}}(\mathbf{y})G^{0^-
}(\mathbf{x},\mathbf{y})+ \int_{-\infty}^{\infty}
A^{0^-}\exp\left(\mbox{i}k_1x_1-\mbox{i}k_2^{0}\left(x_2-b
\right)\right)\frac{dk_1}{k_2^0} \label{section123e3}
\end{equation}
wherein $\mbox{H}_{\Omega_{j}}$ is the Heaviside function
\begin{equation}
\displaystyle \mbox{H}_{\Omega_{j}}(\mathbf{y})=\left\{
\begin{array}{ll}
\displaystyle 1 & \displaystyle \mbox{if } \mathbf{y} \in \Omega_j \\[8pt]
\displaystyle 0 & \displaystyle \mbox{if } \mathbf{y} \in \Omega_i \mbox{, }i\neq j
\end{array}
\right.
\label{section123e4}
\end{equation}
%%
%--------------------------------------------------------------------
\subsubsection{Application of the transmission conditions}\label{section124}
In cartesian coordinates and on account of the orientation of the
two faces of the layer:
\begin{equation}
\boldsymbol{\nu}(\mathbf{x})\cdot\nabla
\mathcal{F}=\frac{\partial}{\partial x_2}\mathcal{F}~.
\label{section124e1}
\end{equation}
After introducing the fields expressions, eqs.(\ref{section123e1}), (\ref{section123e2}) and (\ref{section123e3}) into the boundary conditions eqs.(\ref{section122e2}) and (\ref{section112e3}), we multiply these relations by $\exp\left(-\mbox{i}K_1 x_1 \right)$ and then integrate from
$-\infty$ to $+\infty$, using the identity
\begin{equation}
\int_{-\infty}^{\infty}\exp\left(\mbox{i}\left(k_1-K_1 \right)x_1 \right) d x_1=
2\pi \delta\left(k_1-K_1 \right)\mbox{, $\delta\left(k_1-K_1 \right)$ being the Kronecker symbol}
\label{section124e9}
\end{equation}
so as to obtain the matrix equation
\begin{equation}
\left(\!
\begin{array}{llll}
\displaystyle 1&\displaystyle \frac{-k_2^0 e^{-\mbox{i}k_2^1a}}{k_2^1}&\displaystyle
\frac{-k_2^0 e^{\mbox{i}k_2^1a}}{k_2^1} &\displaystyle 0\\[8pt]
\displaystyle 1&\displaystyle \frac{\rho^0 e^{-\mbox{i}k_2^1a}}{\rho^1}&\displaystyle
\frac{-\rho^0 e^{\mbox{i}k_2^1a}}{\rho^1} &\displaystyle 0\\[8pt]
\displaystyle 0&\displaystyle \frac{-k_2^0 e^{-\mbox{i}k_2^1b}}{k_2^1}&\displaystyle
\frac{-k_2^0 e^{\mbox{i}k_2^1b}}{k_2^1} &\displaystyle 1\\[8pt]
\displaystyle 0&\displaystyle \frac{-\rho^0 e^{-\mbox{i}k_2^1b}}{\rho^1}&\displaystyle
\frac{\rho^0 e^{\mbox{i}k_2^1a}}{\rho^1} &\displaystyle 1
\end{array}\!
\right)\!
\left(\!
\begin{array}{l}
\displaystyle B^{0^+}\\[8pt]
\displaystyle A^{1}\\[8pt]
\displaystyle B^{1}\\[8pt]
\displaystyle A^{0^-}\\[8pt]
\end{array}\!
\right)=\frac{\mbox{i}e^{-\mbox{i}k_1y_1}}{4\pi}
\left(\!
\begin{array}{l}
\displaystyle -e^{\mbox{i}k_2^0\left(y_2-a\right)}\mbox{H}_{\Omega_{0^+}}-
\frac{k_2^0}{k_2^1}e^{\mbox{i}k_2^1\left(a-y_2\right)}\mbox{H}_{\Omega_{1}}\\[8pt]
\displaystyle e^{\mbox{i}k_2^0\left(y_2-a\right)}\mbox{H}_{\Omega_{0^+}}-
\frac{\rho^0}{\rho^1}e^{\mbox{i}k_2^1\left(a-y_2\right)}\mbox{H}_{\Omega_{1}}\\[8pt]
\displaystyle -e^{\mbox{i}k_2^0\left(b-y_2\right)}\mbox{H}_{\Omega_{0^-}}+
\frac{k_2^0}{k_2^1}e^{\mbox{i}k_2^1\left(y_2-b\right)}\mbox{H}_{\Omega_{1}}\\[8pt]
\displaystyle e^{\mbox{i}k_2^0\left(b-y_2\right)}\mbox{H}_{\Omega_{0^-}}+
\frac{\rho^0}{\rho^1}e^{\mbox{i}k_2^1\left(y_2-b\right)}\mbox{H}_{\Omega_{1}}\\[8pt]
\end{array}
\!
\right)
\label{section124e10}
\end{equation}
%%%%%%%%%%%%%%%%%%%%%%%%%%%%%%%%%%%%%%%%%%%%%%%%%%%%%%%%%%%%%%%%%%%%%
\subsubsection{Final expressions of the specific Green's function}\label{section125}
Once the matrix system (\ref{section124e10}) is solved for
$B^{0^+}$, $A^{1}$, $B^{1}$ and $A^{0^-}$, and  these expressions
are introduced into the  expressions  of the fields
(\ref{section123e1}), (\ref{section123e2}), (\ref{section123e3}),
we get:
\begin{multline}
g^{0^+}(\mathbf{x},\mathbf{y})=\frac{\mbox{i}}{4
\pi}\int_{-\infty}^{\infty} e^{[\mbox{i}k_1\left(x_1-y_1\right)+
\mbox{i}k_2^0|x_2-y_2|]}\mbox{H}_{\Omega_{0^+}}\frac{dk_1}{k_2^0}+
\\
\int_{-\infty}^{\infty}\frac{e^{[\mbox{i}k_1\left(x_1-y_1\right)+
\mbox{i}k_2^{0}x_2]}}{4 \pi \left(2\alpha^0\alpha^1\cos\left(k_2^1
l \right)- \mbox{i}\left((\alpha^0)^2+(\alpha^1)^2
\right)\sin\left(k_2^1 l \right) \right)}\times\\
 \left[e^{\mbox{i}k_2^0\left(y2-2a
\right)}\sin\left(k_2^1 l \right)\left((\alpha^0)^2-(\alpha^1)^2
\right)\frac{\mbox{H}_{\Omega_{0^+}}}{k_2^0}+2\mbox{i}
e^{-\mbox{i}k_2^0\left(y_2+l
\right)}\alpha^1\alpha^0\frac{\mbox{H}_{\Omega_{0^-}}}{k_2^0}+\right.
\\
\left.2\mbox{i}e^{-\mbox{i}k_2^0a}\alpha^1\left(\alpha^1\cos\left(k_2^1\left(y_2-b\right)
\right)-\mbox{i}\alpha^0\sin\left(k_2^1\left(y_2-b\right) \right)
\right) \frac{\mbox{H}_{\Omega_{1}}}{k_2^1}\right]dk_1 ~,
\label{section125e1}
\end{multline}
\begin{multline}
\displaystyle g^{1}(\mathbf{x},\mathbf{y})=\frac{\mbox{i}}{4
\pi}\int_{-\infty}^{\infty}
e^{[\mbox{i}k_1\left(x_1-y_1\right)+\mbox{i}k_2^1|x_2-y_2|]}\mbox{H}_{\Omega_1}\frac{dk_1}{k_2^1}+
\\
\int_{-\infty}^{\infty}\frac{\mbox{i}e^{\mbox{i}k_1\left(x_1-y_1\right)}}{4
\pi \left(2\alpha^0\alpha^1\cos\left(k_2^1 l
\right)-\mbox{i}\left((\alpha^0)^2+(\alpha^1)^2
\right)\sin\left(k_2^1 l \right) \right)}\times
\\
\left[2e^{\mbox{i}k_2^0\left(y_2-a
\right)}\left(\alpha^1\alpha^0\cos\left(k_2^1 \left(x_2-b\right)
\right)-\mbox{i}\left(\alpha^0\right)^2\sin \left(k_2^1
\left(x_2-b\right) \right)
\right)\frac{\mbox{H}_{\Omega_{0^+}}}{k_2^0}+\right.
\\
 2e^{\mbox{i}k_2^0\left(b-y_2
\right)}\left(\alpha^1\alpha^0\cos\left(k_2^1 \left(a-x_2\right)
\right)-\mbox{i}\left(\alpha^0\right)^2\sin \left(k_2^1
\left(a-x_2\right) \right)
\right)\frac{\mbox{H}_{\Omega_{0^-}}}{k_2^0}+
\\
\left.\left(\left((\alpha^1)^2-(\alpha^0)^2\right)\cos\left(k_2^1\left(x_2+y_2-a-b\right)
\right)+\exp\left(\mbox{i}k_{2}^{1}l
\right)\left(\alpha^0-\alpha^1\right)^2\cos\left(k_2^1\left(x_2-y_2\right)
\right) \right) \frac{\mbox{H}_{\Omega_{1}}}{k_2^1}\right]dk_1 ~,
\label{section125e2}
\end{multline}
\begin{multline}
g^{0^-}(\mathbf{x},\mathbf{y})= \frac{\mbox{i}}{4
\pi}\int_{-\infty}^{\infty}e^{[\mbox{i}k_1\left(x_1-y_1\right)+\mbox{i}k_2^0|x_2-y_2|]}
\mbox{H}_{\Omega_{0^-}}\frac{dk_1}{k_2^0}+
\\
\int_{-\infty}^{\infty}\frac{e^{[\mbox{i}k_1\left(x_1-y_1\right)-\mbox{i}k_2^{0}x_2]}}{4
\pi \left(2\alpha^0\alpha^1\cos\left(k_2^1 l
\right)-\mbox{i}\left((\alpha^0)^2+(\alpha^1)^2
\right)\sin\left(k_2^1 l \right) \right)}\times
\\
\left[2\mbox{i}e^{-\mbox{i}k_2^0\left(l-y_2
\right)}\alpha^1\alpha^0\frac{\mbox{H}_{\Omega_{0^+}}}{k_2^0}+
e^{-\mbox{i}k_2^0\left(y2-2b \right)}\sin\left(k_2^1 l
\right)\left((\alpha^0)^2-(\alpha^1)^2
\right)\frac{\mbox{H}_{\Omega_{0^-}}}{k_2^0}+\right.
\\
\left.2\mbox{i}e^{\mbox{i}k_2^0b}\alpha^1\left(\alpha^1\cos\left(k_2^1\left(a-y_2\right)
\right)-\mbox{i}\alpha^0\sin\left(k_2^1\left(a-y_2\right) \right)
\right) \frac{\mbox{H}_{\Omega_{1}}}{k_2^1}\right]dk_1 ~.
\label{section125e3}
\end{multline}
%%
%%%%%%%%%%%%%%%%%%%%%%%%%%%%%%%%%%%%%%%%%%%%%%%%%%%%%%%%%%%%%%%%%%%%%
\subsection{Use of a specific Green's function to solve the
direct problem of pressure wave
scattered by an inhomogeneous fluid-filled slab}\label{section13}
\begin{figure}[H]
\centering\psfig{figure=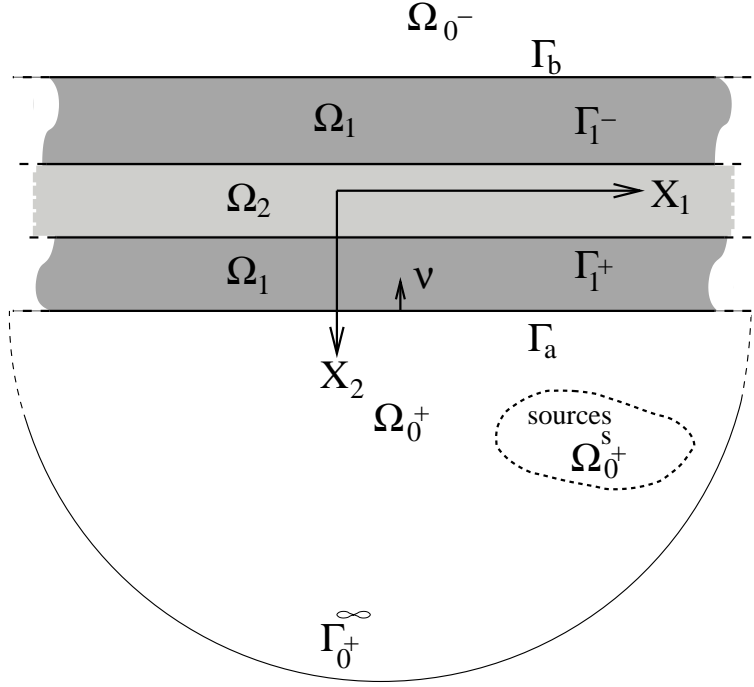,width=10cm}
\caption{Configuration of the direct problem of a fluid heterogeneity
within a fluid-like slab.}
\label{section13figure1}
\end{figure}

We treat the 2D fluid acoustic direct problem illustrated in
figure \ref{section13figure1}. In the absence of the heterogeneity,
occupying the domain $\Omega_2$, the configuration is that of the
closed layer domain $\Omega_1$ occupied by a known homogeneous
fluid $M^1$ with (spatially-constant) acoustic parameters ($k^1$,
$\rho^1$), surrounded by the open domain $\Omega_0$ occupied by a
known homogeneous fluid $M^0$ with (spatially-constant) acoustic
parameters ($k^0$, $\rho^0$).

In the presence of the heterogeneity, localized to the domain
$\Omega_2\in\Omega_1$, the problem is to solve the scattering
problem for spatially-varying acoustic parameter functions
($k^2(\mathbf{x})$, $\rho^2(\mathbf{x})$) of the medium $M^2$
filling $\Omega_2$  in the subdomains $\Omega_{0^+}$ and
$\Omega_{0^-}$ when the slab is probed by an incident wave.
%%
%--------------------------------------------------------------------
\subsubsection{Governing equations for scattering from a heterogeneous layer, included
between $\Gamma_a$ and $\Gamma_b$, probed by a cylindrical wave
radiated by a cylindrical source whose support is
$\Omega_{0^+}^{s}$}\label{section131}
 Let
$\Omega_1=\widetilde{\Omega}_1\,\cup\, \Omega_2$. Then the
governing equations for the pressure field are:
\begin{equation}
\left[\triangle +(k^0)^2 \right] p^{0^+}(\mathbf{x})=
-s^0(\mathbf{x})\mbox{; }\mathbf{x}\in \Omega_{0^+},
\label{section131e1}
\end{equation}
\begin{equation}
\left[\triangle +\left(k(\mathbf{x})\right)^2 \right]
p^1(\mathbf{x})= \frac{\nabla
\rho(\mathbf{x})}{\rho(\mathbf{x})}\cdot \nabla
p^1(\mathbf{x})\mbox{; }\mathbf{x}\in \Omega_1,
\label{section131e2}
\end{equation}
\begin{equation}
k(\mathbf{x})=
\left\{
\begin{array}{lll}
\displaystyle k^1&\displaystyle ;&\displaystyle \mathbf{x}\in \widetilde{\Omega}_1\\[8pt]
\displaystyle k^2(\mathbf{x})&\displaystyle ;&\displaystyle \mathbf{x}\in \Omega_2
\end{array}
\right.
\label{section131e3}
\end{equation}
\begin{equation}
\rho(\mathbf{x})=
\left\{
\begin{array}{lll}
\displaystyle \rho^1&\displaystyle ;&\displaystyle \mathbf{x}\in \widetilde{\Omega}_1\\[8pt]
\displaystyle \rho^2(\mathbf{x})&\displaystyle ;&\displaystyle \mathbf{x}\in \Omega_2
\end{array}
\right.
\label{section131e3bis}
\end{equation}
\begin{equation}
\left[\triangle +(k^0)^2 \right] p^{0^-}(\mathbf{x})=0\mbox{;
}\mathbf{x}\in \Omega_{0^-}, \label{section131e4}
\end{equation}
\begin{equation}
p^{0+}(\mathbf{x})-p^1(\mathbf{x})=0\mbox{; }\mathbf{x}\in
\Gamma_{a}, \label{section131e5}
\end{equation}
\begin{equation}
\frac{1}{\rho^0} \boldsymbol{\nu}(\mathbf{x})\cdot \nabla
p^{0+}(\mathbf{x})- \frac{1}{\rho^1}
\boldsymbol{\nu}(\mathbf{x})\cdot \nabla p^1(\mathbf{x})=0\mbox{;
}\mathbf{x}\in \Gamma_{a}, \label{section131e6}
\end{equation}
\begin{equation}
p^{1}(\mathbf{x})-p^{0^-}(\mathbf{x})=0\mbox{; }\mathbf{x}\in
\Gamma_{b}, \label{section131e7}
\end{equation}
\begin{equation}
\frac{1}{\rho^1} \boldsymbol{\nu}(\mathbf{x})\cdot \nabla
p^{1}(\mathbf{x})- \frac{1}{\rho^0}
\boldsymbol{\nu}(\mathbf{x})\cdot \nabla
p^{0^-}(\mathbf{x})=0\mbox{; }\mathbf{x}\in \Gamma_{b},
\label{section131e8}
\end{equation}
\begin{equation}
p^{0^+}(\mathbf{x})\mbox{, }p^{1}(\mathbf{x})\mbox{ and }
p^{0^-}(\mathbf{x})\sim\mbox{ outgoing waves,
}\left\|\mathbf{x}\right\| \rightarrow \infty \label{section131e9}
\end{equation}
%%
%--------------------------------------------------------------------
%\subsubsection{Governing equations for the specific Green's function}\label{section132}
The previously-given governing equations for the specific Green's
function can be rewritten as:
\begin{equation}
\left[\triangle +(K(\mathbf{x}))^2 \right]g
(\mathbf{x},\mathbf{y})= -\delta(\mathbf{x}-\mathbf{y})\mbox{;
}\mathbf{x}\in \mathbb{R}^2\mbox{, }\mathbf{y}\in \mathbb{R}^2
\label{section132e1}
\end{equation}
\begin{equation}
K(\mathbf{x})=
\left\{
\begin{array}{lll}
\displaystyle k^0&\displaystyle ;&\displaystyle \mathbf{x}\in \Omega_{0^+}\, \cup\, \Omega_{0^-}\\[8pt]
\displaystyle k^1&\displaystyle ;&\displaystyle \mathbf{x}\in \Omega_1
\end{array}
\right.
\label{section132e2}
\end{equation}
\begin{equation}
g(\mathbf{x},\mathbf{y})= \left\{
\begin{array}{lll}
\displaystyle g^0(\mathbf{x},\mathbf{y})&\displaystyle
;&\displaystyle \mathbf{x}\in \Omega_{0^+}\, \cup\,
\Omega_{0^-}\\[8pt] \displaystyle
g^1(\mathbf{x},\mathbf{y})&\displaystyle ;&\displaystyle
\mathbf{x}\in \Omega_1
\end{array}
\right. \label{section132e2a}
\end{equation}
\begin{equation}
g^{0+}(\mathbf{x},\mathbf{y})-g^1(\mathbf{x},\mathbf{y})=0\mbox{;
}\mathbf{x}\in \Gamma_{a}, \label{section132e3}
\end{equation}
\begin{equation}
\frac{1}{\rho^0} \boldsymbol{\nu}(\mathbf{x})\cdot \nabla
g^{0+}(\mathbf{x},\mathbf{y})- \frac{1}{\rho^1}
\boldsymbol{\nu}(\mathbf{x})\cdot \nabla
g^1(\mathbf{x},\mathbf{y})=0\mbox{; }\mathbf{x}\in \Gamma_{a},
\label{section132e4}
\end{equation}
\begin{equation}
g^{1}(\mathbf{x},\mathbf{y})-g^{0^-}(\mathbf{x},\mathbf{y})=0\mbox{;
}\mathbf{x}\in \Gamma_{b}, \label{section132e5}
\end{equation}
\begin{equation}
\frac{1}{\rho^1} \boldsymbol{\nu}(\mathbf{x})\cdot \nabla
g^{1}(\mathbf{x},\mathbf{y})- \frac{1}{\rho^0}
\boldsymbol{\nu}(\mathbf{x})\cdot \nabla
g^{0^-}(\mathbf{x},\mathbf{y})= 0\mbox{; }\mathbf{x}\in
\Gamma_{b}, \label{section132e6}
\end{equation}
\begin{equation}
g(\mathbf{x},\mathbf{y})\sim\mbox{ outgoing waves,
}\left\|\mathbf{x}\right\| \rightarrow \infty \label{section132e9}
\end{equation}
%%
%%
%--------------------------------------------------------------------
\subsubsection{Towards a domain integral representation of the pressure field
in $\Omega_{0^+}$}\label{section133}
In obvious short-hand notation (in addition:
$\partial_{\nu}:=\boldsymbol{\nu}\cdot \nabla$), we obtain from
the previous governing equations:
\begin{equation}
g^{0^+}\left[\triangle +(k^0)^2 \right] p^{0^+}=-g^{0^+} s^0\mbox{ ; in } \Omega_{0^+}
\label{section133e1}
\end{equation}
\begin{equation}
p^{0^+}\left[\triangle +(k^0)^2 \right] g^{0^+}=-p^{0^+} \delta\mbox{ ; in } \Omega_{0^+}
\label{section133e2}
\end{equation}
so that integrating  the difference of these two equations over
$\Omega_{0^+}$, we obtain
\begin{equation}
\int_{\Omega_{0}} \left(g^{0^+} \triangle p^{0^+}-p^{0^+} \triangle g^{0^+} \right)d \Omega=
-\int_{\Omega_{0}}g^{0^+} s^0d \Omega+\int_{\Omega_{0}}p^{0^+} \delta d \Omega
\label{section133e3}
\end{equation}
or, after use of Green's theorem and the sifting property of the
$\delta$ distribution:
\begin{equation}
\begin{array}{l}
\displaystyle \int_{\Gamma_{0^+}^{\infty}} \left(g^{0^+} \partial_{\nu} p^{0^+}-p^{0^+} \partial_{\nu} g^{0^+} \right)d \gamma+\\[8pt]
\displaystyle \int_{\Gamma_{a}} \left(g^{0^+} \partial_{\nu} p^{0^+}-p^{0^+} \partial_{\nu} g^{0^+} \right)d \gamma+
\int_{\Omega_{0}}g^{0^+} s^0d \Omega=p^{0^+}(\mathbf{y})\mbox{H}_{\Omega^{0^+}}(\mathbf{y})
\end{array}
\label{section133e4}
\end{equation}
We develop, for the domain integral representation of this pressure field, the integration over $\Gamma_{0^+}^{\infty}$, so that to obtain:
\begin{equation}
\int_{\Gamma_{0^+}^{\infty}} \left(g^{0^+} \partial_{\nu} p^{0^+}-p^{0^+} \partial_{\nu} g^{0^+} \right)d \gamma=\int_{\Gamma_{0^+}^{\infty}} g^{0^+} \left[\partial_{\nu} p^{0^+}-\mbox{i} k^0p^{0^+} \right]d \gamma - \int_{\Gamma_{0^+}^{\infty}}p^{0^+} \left[\partial_{\nu} g^{0^+}-\mbox{i} k^0 g^{0^+}  \right]d \gamma
\label{section133e5}
\end{equation}
It is readily shown that both of the integrals on the right hand side of this expression vanish due to the fact that both $p^{0^+}$ and $g^{0^+}$ satisfy the (frequency domain) rediation condition at infinity.
%%
%--------------------------------------------------------------------
\subsubsection{Towards a domain integral representation of the pressure
field in $\Omega_{1}$}\label{section134}
In obvious short-hand notation, we obtain from the previous
governing equations:
\begin{equation}
\left[\triangle +(k(\mathbf{x}))^2 \right] p^{1}= \frac{\nabla
\rho}{\rho}\cdot \nabla p^1 \mbox{ ; }\Rightarrow \left[\triangle
+(k^1)^2 \right] p^{1}= \left[(k^{1})^2-(k(x))^2\right]
p^1+\frac{\nabla \rho}{\rho}\cdot \nabla p^1=-\sigma(\mathbf{x})
\label{section134e1}
\end{equation}
Consequently,
\begin{equation}
g^{1}\left[\triangle +(k^1)^2 \right] p^{1}=-g^{1}\sigma \delta\mbox{ ; in } \Omega_{1}
\label{section134e2}
\end{equation}
\begin{equation}
p^{1}\left[\triangle +(k^1)^2 \right] g^{1}=-p^{1} \delta\mbox{ ; in } \Omega_{1}
\label{section134e3}
\end{equation}
so that, integrating  the difference of these two equations over
$\Omega_{1}$, and after use of Green's theorem and the sifting property of the
$\delta$ distribution, we obtain:
%%
%\begin{equation}
%\int_{\Omega_{1}} \left(g^{1} \triangle p^{1}-p^{1} \triangle g^{1} \right)d \Omega=
%-\int_{\Omega_{1}}g^{1} \sigma \Omega+\int_{\Omega_{1}}p^{1} \delta d \Omega
%\label{section134e4}
%\end{equation}
%%
%or, :
%%
\begin{equation}
-\int_{\Gamma_{a}} \left(g^{1} \partial_{\nu} p^{1}-
p^{1} \partial_{\nu} g^{1} \right) d \gamma+\int_{\Gamma_{b}} \left(g^{1} \partial_{\nu} p^{1}-
p^{1} \partial_{\nu} g^{1} \right) d \gamma +\int_{\Omega_{1}} g^{1} \sigma d \Omega=
p^{1}(\mathbf{y})\mbox{H}_{\Omega^{1}}(\mathbf{y})
\label{section134e5}
\end{equation}
which yields, on account of the transmission conditions:
\begin{multline}
-\int_{\Gamma_{a}} \left(g^{0^+} \partial_{\nu} p^{0^+}-
p^{0^+} \partial_{\nu} g^{0^+} \right) d \gamma+
\int_{\Gamma_{b}} \left(g^{0^-} \partial_{\nu} p^{0^-}-
p^{0^-} \partial_{\nu} g^{0^-} \right) d \gamma +
\frac{\rho^0}{\rho^1}\int_{\Omega_{1}} g^{1} \sigma d \Omega=
\\
\frac{\rho^0}{\rho^1}p^{1}(\mathbf{y})\mbox{H}_{\Omega^{1}}(\mathbf{y})
\label{section134e6}
\end{multline}
%%
%--------------------------------------------------------------------
\subsubsection{Towards a domain integral representation of the pressure
field in $\Omega_{0^-}$}\label{section135}
In obvious short-hand notation, we obtain from the previous
governing equations:
\begin{equation}
g^{0^-}\left[\triangle +(k^0)^2 \right] p^{0^-}=0\mbox{ ; in }
\Omega_{0^-} ~,
\label{section135e1}
\end{equation}
\begin{equation}
p^{0^-}\left[\triangle +(k^0)^2 \right] g^{0^-}=-p^{0^-}
\delta\mbox{ ; in } \Omega_{0^-} ~,
\label{section135e2}
\end{equation}
so that, integrating the difference of these two equations over
$\Omega_{0^-}$, and following the procedure used in the last two subsections, we obtain:
\begin{equation}
-\int_{\Gamma_{b}} \left(g^{0^-} \partial_{\nu} p^{0^-}-
p^{0^-} \partial_{\nu} g^{0^-} \right)d \gamma=
p^{0^-}(\mathbf{y})\mbox{H}_{\Omega^{0^-}}(\mathbf{y})
 ~.
\label{section135e4}
\end{equation}
%%
%--------------------------------------------------------------------
\subsubsection{Domain integral representations, without boundary terms, of the pressure
fields in $\Omega_{0^+}$, $\Omega_{1}$ and $\Omega_{0^-}$}\label{section136}
The addition of (\ref{section133e4}), (\ref{section134e6}) and (\ref{section135e4}) gives
\begin{equation}
\int_{\Omega_{0^+}} g^{0^+}s^0 d\Omega+\frac{\rho^0}{\rho^1}\int_{\Omega_{1}} g^{1}\sigma d\Omega=
p^{0^+}(\mathbf{y})\mbox{H}_{\Omega^{0^+}}(\mathbf{y})+
p^{1}(\mathbf{y})\mbox{H}_{\Omega^{1}}(\mathbf{y})+p^{0^+}(\mathbf{y})\mbox{H}_{\Omega^{0^+}}(\mathbf{y})
 ~,
\label{section136e1}
\end{equation}
from which it ensues, on account of the properties of the domain Heaviside function:
\begin{equation}
p^{0^+}(\mathbf{y})=\int_{\Omega_{0^+}} g^{0^+}s^0
d\Omega+\frac{\rho^0}{\rho^1}\int_{\Omega_{1}}g^{1}(\mathbf{x},\mathbf{y})
\left[\left(k(\mathbf{x})^2-(k^1)^2\right)-\frac{\nabla
\rho}{\rho}\cdot \nabla \right]p^1(\mathbf{x})d\Omega\mbox{,
}\forall \mathbf{y} \in \Omega_{0^+}  ~,
\label{section136e2}
\end{equation}
\begin{equation}
p^{1}(\mathbf{y})=\frac{\rho^1}{\rho^0}\int_{\Omega_{0^+}}
g^{0^+}s^0
d\Omega+\int_{\Omega_{1}}g^{1}(\mathbf{x},\mathbf{y})\left[\left(k(\mathbf{x})^2-(k^1)^2\right)-\frac{\nabla
\rho}{\rho}\cdot \nabla \right]p^1(\mathbf{x})d\Omega\mbox{,
}\forall \mathbf{y} \in \Omega_{1} ~,
\label{section136e3}
\end{equation}
\begin{equation}
p^{0^-}(\mathbf{y})=\int_{\Omega_{0^+}} g^{0^+}s^0
d\Omega+\frac{\rho^0}{\rho^1}\int_{\Omega_{1}}g^{1}(\mathbf{x},\mathbf{y})
\left[\left(k(\mathbf{x})^2-(k^1)^2\right)-\frac{\nabla
\rho}{\rho}\cdot \nabla \right]p^1(\mathbf{x})d\Omega\mbox{,
}\forall \mathbf{y} \in \Omega_{0^-} ~.
\label{section136e4}
\end{equation}
%%
%%%%%%%%%%%%%%%%%%%%%%%%%%%%%%%%%%%%%%%%%%%%%%%%%%%%%%%%%%%%%%%%%%%%%
\subsubsection{Other integral representations, without boundary terms,
of the pressure fields in $\Omega_{0^+}$, $\Omega_{1}$ and $\Omega_{0^-}$}
\label{section138}
Let $g_i^j(\mathbf{x},\mathbf{y})$ correspond to
$\mathbf{y}\in\Omega_{i}$ and $\mathbf{x}\in\Omega_{j}$.
Reciprocity implies $\displaystyle
g_i^j(\mathbf{x},\mathbf{y})=\frac{\rho^j}{\rho^i}g_j^i(\mathbf{y},\mathbf{x})$.
The integral representations (\ref{section136e2}),
(\ref{section136e3}) and (\ref{section136e4})
 can finally be written in the
condensed form (recalling that $\Omega=\Omega_{0^+} \cup
\Omega_{1} \cup\Omega_{0^-}$):
\begin{equation}
\displaystyle p(\mathbf{y},\omega)=\int_{\Omega_{0^+}^s}
g_{0^+}(\mathbf{y},\mathbf{x})s^0(\mathbf{x}) d\Omega(\mathbf{x})+
\int_{\Omega_{1}}g_{1}(\mathbf{y},\mathbf{x})\left[\left(k(\mathbf{x})^2-
(k^1)^2\right)-\frac{\nabla \rho}{\rho}\cdot \nabla \right]
p(\mathbf{x})d\Omega(\mathbf{x})\mbox{; }\forall \mathbf{y} \in
\Omega \label{section138e4}
\end{equation}
%%
%%%%%%%%%%%%%%%%%%%%%%%%%%%%%%%%%%%%%%%%%%%%%%%%%%%%%%%%%%%%%%%%%%%%%
\section{Comments on the integral representation of the field}\label{section2}
Eq. (\ref{section138e4}) can be written as:
\begin{multline}
p^{s}(\mathbf{y}):=p(\mathbf{y})-\int_{\Omega_{0^+}^s}
g_{0^+}(\mathbf{y},\mathbf{x})s^0(\mathbf{x}) d\Omega(\mathbf{x})=
\\
\int_{\Omega_{1}}g_{1}(\mathbf{y},\mathbf{x})\left[\left(k(\mathbf{x})^2-(k^1)^2\right)-\frac{\nabla
\rho}{\rho}\cdot \nabla
\right]p(\mathbf{x})d\Omega(\mathbf{x})\mbox{; } \forall
\mathbf{y} \in \Omega ~,
\label{section2e4}
\end{multline}
wherein $p^{s}(\mathbf{y})$ is the field scattered  by the
inhomogeneity of the slab.

This can be compared to the more common formulation employing
the free-space Green's function ($G^0(\mathbf{x},\mathbf{y})$):
\begin{multline}
p^{d}(\mathbf{y}):= p(\mathbf{y},\omega)-\int_{\Omega_{0^+}^s}
G^{0}(\mathbf{y},\mathbf{x})s^0(\mathbf{x}) d\Omega(\mathbf{x})=
\\
\int_{\Omega_{1}}G^0(\mathbf{y},\mathbf{x})\left[(k(\mathbf{x}))^2-(k^0)^2-
\frac{\nabla \rho}{\rho}\cdot \nabla
\right]p(\mathbf{x})d\Omega(\mathbf{x})\mbox{; } \forall
\mathbf{y} \in \Omega ~, \label{section2e5}
\end{multline}
wherein $p^{d}(\mathbf{y},\omega)$ is the field diffracted  by the
entire inhomogeneous slab (including the slab itself and its
inhomogeneities). The SGF formulation thus appears to be more
suitable than the FSGF formulation, because the scattered field
$p^{s}(\mathbf{y},\omega)$ accounts at the outset for more of the physics of the
interaction of the obstacle with the incident wave
than $p^{d}(\mathbf{y})$.

%Ajount2
It can be shown that the neglected field  is generally smaller in
the SGF formulation than in the FSFG formulation. Effectively,
when the SGF formulation is employed, the zeroth order Born
approximation  consists in neglecting $\displaystyle
\int_{\Omega_{1}}g_{1}(\mathbf{y},\mathbf{x})
\left[\left(k(\mathbf{x})^2-(k^1)^2\right)-\frac{\nabla
\rho}{\rho}\cdot \nabla \right]p(\mathbf{x})d\Omega(\mathbf{x})$
compared to $\displaystyle \int_{\Omega_{0^+}^s}
g_{0^+}(\mathbf{y},\mathbf{x})s^0(\mathbf{x}) d\Omega(\mathbf{x})$
whereas when  the FSFG formulation is employed,  the zeroth-order
Born approximation consists in neglecting
$\int_{\Omega_{1}}G^0(\mathbf{y},\mathbf{x})
\left[(k(\mathbf{x}))^2-(k^0)^2-\frac{\nabla \rho}{\rho}\cdot
\nabla \right]p(\mathbf{x})d\Omega(\mathbf{x})$ in comparison to
$\int_{\Omega_{0^+}^s} G^{0}(\mathbf{y},\mathbf{x})s^0(\mathbf{x})
d\Omega(\mathbf{x}) $.

The use of the SGF includes some multiple reflections, while the
FSGF formulation combined with the Born approximation does not
apply to high contrasts, because the employed linearization
tacitly precludes multiple reflections.

Finally, when the density is constant, the contrast component of the
kernels of both formulations reduce to
\begin{equation}
(k^{j})^2\left(\left(\frac{k(x_2)}{k^j} \right)^2-1 \right)=
(k^{j})^2\left(\left(\frac{\frac{\omega}{c(x_2)}+\mbox{i}
\alpha(\omega,x_2)}{k^j} \right)^2-1 \right)
\label{section2e6}
\end{equation}
wherein $\alpha(\omega,x_2)$ is the absorption coefficient and
$j=0$ for the FSGF and $j=1$ for SGF. It has been shown in
\cite{kakandslaney} that the Born approximation is reasonable if
the phase shift introduced by the inhomogeneous medium is less
than $\pi$, i.e., weak and smooth heterogeneities of simple shape.
The shift depends not only on the size, but also on the kernel,
(eq. \ref{section2e6}), i.e., on the frequency, on the absorption,
and on the contrast between the two phase velocities. The use of
the SGF, when the initial configuration is a homogeneous slab
filled with a fluid-saturated porous material, allows us: i) to
reduce the frequency dependence of the kernel, ii) to reduce the
kernel itself by taking into account a phase-velocity that is
closer to that of the host medium and also by taking into account
the absorption (dissipation) of the material, and iii) to provide
more accuracy, in the sense that on the one hand, the specific
Green's function  already accounts  for dissipation and for some
of the geometry of the problem and, on the other hand, the
approximation of the field in the integral is more realistic than
when the FSFG is employed. The usual way (i.e. when the FSFG is
used) to avoid the problem induced by the absorption  consists in
adding some dissipation term in the approximated field in the
integrand (Modified Born approximation), but not by acting
directly on the kernel of the integral. Methods such as the
distorted Born approximation, whose convergence analysis has been
carried out in \cite{tijhuis}, also allow to consider objects with
larger contrast, but by acting only on the constrast function,
i.e.,  without introducing additional effects on the approximated
field in the integrand and on the Green's function used in the
formulation.

The SGF domain integral formulation thus allows the elimination of
some of the disavantages of the FSGF domain integral formulation.
This is obtained  by acting on the kernel, the Green's function
and the approximated pressure field, contrary to other methods
employing the FSFG which act only on one or two of the components
of the integrand.

The combined effects of this action is to allow us to define and
implement an iterative scheme, starting with the zeroth-order Born
approximation and using the SGF formulation, to solve wave
propagation problems involving a medium, whose components have high
constrasts and in which there exist abrupt heterogeneities, so
to consider objects with larger constrats, with respect to the
surrounding medium, than would be possible with the conventional
FSGF formulation.
%Fin Ajout 2
%%%%%%%%%%%%%%%%%%%%%%%%%%%%%%%%%%%%%%%%%%%%%%%%%%%%%%%%%%%%%%%%%%%%%
\section{Specific ingredients of the computational procedure for
the prediction of the field scattered  by an inhomogeneous porous
slab solicited by a plane incident wave.}\label{section3}
We now adapt the previous analysis to the determination of the
field scattered by an inhomogeneous slab (the direction of
inhomogeneity being $x_2$) solicited by an incident plane wave.

This type of  incident wave is associated with  $s^0=0$, so that
it would appear that there is no solicitation in the above
equations. Nevertheless, for an incident plane wave initially
propagating in $\Omega_{0^+}$, the integral over
$\Gamma_{0^+}^{\infty}$, (\ref{section133e5}), does not vanish. It
follows that the term corresponding to the solicitation takes the
form of $p^{0^+}(\mathbf{y})$, $p^{1}(\mathbf{y})$ and
$p^{0^-}(\mathbf{y})$, which are the responses in the subdomains
$\Omega_{0^+}$, $\Omega_{1}$ and  $\Omega_{0^-}$ respectively
(i.e. the zeroth-order Born approximation) to an incident plane
wave propagating initially in $\Omega^{0^+}$ given by
\begin{equation}
p^{i}(\mathbf{y},\omega)=A^{i}(\omega)\exp[\mbox{i}(k_{1}^{i}x_{1}-k_{2}^{0,i}x_{2})]
\label{section3e1a}
\end{equation}
wherein $k_{1}^{i}=k^{0}\sin\theta^{i}$,
$k_{2}^{0,i}=k^{0}\cos\theta^{i}$ and $\theta^{i}$ the angle of
incidence with respect to the $+x_{2}$ axis.
The spectrum of the incident takes the form of a Ricker-like wavelet of the form:
\begin{equation}
A^{i}(\omega)=\frac{-\left(\pi \nu_0 \right)^{2} \omega^{2}}{2
\sqrt{\pi}\left(\pi \nu_0 \right)^{3}}\exp\left(\frac{\mbox{i}
\omega}{\nu_0}-\frac{\omega^2}{\left(2 \pi \nu_0 \right)^{2}}
\right) \label{section3e2}
\end{equation}
wherein we take (in the computations) $\nu_0=100kHz$ to be the
central frequency of the source spectrum.

The zeroth-order Born approximation is given in appendix
\ref{appendixe1}.
%Fin ajout2
%%%%%%%%%%%%%%%%%%%%%%%%%%%%%%%%%%%%%%%%%%%%%%%%%%%%%%%%%%%%%%%%%%%%%
\subsection{Application of the first-order Born approximation in
the SGF formulation}\label{section31}
We give here the explicit form of the first-order Born
approximation within the framework of the SGF formulation.

{\textbf{Remark:}} As a consequence of the separation of
variables, all pressure fields can be written in the form
\begin{equation}
p(\mathbf{x})=\exp\left(\mbox{i}k_1^{i}x_1 \right)\tilde{p}(x_2)
\label{section31e1}
\end{equation}
%%%%%%%%%%%%%%%%%%%%%%%%%%%%%%%%%%%%%%%%%%%%%%%%%%%%%%%%%%%%%%%%%%%%%%
\subsubsection{First-order Born approximation in the SGF formulation
for $\mathbf{y}\in \Omega_{0+}$}\label{section311}
When  $\mathbf{y}\in \Omega_{0+}$,
$g_{1}(\mathbf{y},\mathbf{x},\omega)=g_{1}^{0^+}(\mathbf{y},\mathbf{x},\omega)$,
so that
\begin{equation}
\begin{array}{l}
\displaystyle p(\mathbf{y},\omega)-p^{0+}(\mathbf{y},\omega)
\approx\int_{-\infty}^{+\infty}\int_{b}^{a}
g_{1}^{0+}(\mathbf{y},\mathbf{x})\left[\left(k(x_2)^2-(k^1)^2\right)-\frac{1}{\rho(x_2)}\frac{\partial
\rho(x_2)}{\partial x_2}\,\frac{\partial }{\partial x_2}
\right]p^{1}(\mathbf{x})\,dx_1\, dx_2\\[10pt]
\displaystyle \approx\int_{b}^{a}
\left[\int_{-\infty}^{\infty}\frac{\mbox{i}
e^{[\mbox{i}k_1y_1+\mbox{i}k_2^{0}\left(y_2-a\right)]}\alpha^1\left(\alpha^1
\cos\left(k_2^1\left(x_2-b\right)
\right)-\mbox{i}\alpha^0\sin\left(k_2^1\left(x_2-b\right) \right)
\right) } {2 \pi  \left(2\alpha^0\alpha^1\cos\left(k_2^1 l
\right)-\mbox{i}\left((\alpha^0)^2+(\alpha^1)^2 \right)
\sin\left(k_2^1 l \right) \right)} \frac{dk_1}{k_2^1} \right]
\times\\[10pt]
\displaystyle
\left[\left(k(x_2)^2-(k^1)^2\right)-\frac{1}{\rho(x_2)}\frac{\partial
\rho(x_2)} {\partial x_2}\,\frac{\partial }{\partial x_2}
\right]\tilde{p}^{1}(x_2) \left[ \int_{-\infty}^{+\infty}
e^{\mbox{i}\left(k_1^i-k_1\right) x_1}\,dx_1\right]\, dx_2 ~.
\end{array}
\label{section311e1}
\end{equation}
By making use of the identity (\ref{section124e9}), (\ref{section311e1}) becomes

\begin{multline}
p(\mathbf{y},\omega)-p^{0+}(\mathbf{y},\omega) \approx
\\
\int_{b}^{a} \left[\frac{\mbox{i}
e^{[\mbox{i}k_1^iy_1+\mbox{i}k_2^{0,i}\left(y_2-a\right)]}\alpha^{1,i}
\left(\alpha^{1,i}\cos\left(k_2^{1,i}\left(x_2-b\right)
\right)-\mbox{i}\alpha^0 \sin\left(k_2^{1,i}\left(x_2-b\right)
\right) \right) }{ k_2^{1,i} \left(2\alpha^{0,i}\alpha^{1,i}
\cos\left(k_2^{1,i} l
\right)-\mbox{i}\left((\alpha^{0,i})^2+(\alpha^{1,i})^2 \right)
\sin\left(k_2^{1,i} l \right) \right)} \right] \times
\\
\left[\left(k(x_2)^2-(k^1)^2\right)-\frac{1}{\rho(x_2)}\frac{\partial
\rho(x_2)} {\partial x_2}\,\frac{\partial }{\partial x_2}
\right]\tilde{p}^{1}(x_2) \, dx_2 ~.
\label{section311e3}
\end{multline}
Introducing the expression of $\tilde{p}^{1}(x_2)$ from (\ref{appendix1e2}), and after expanding, we get:
\begin{multline}
p(\mathbf{y},\omega)-p^{0+}(\mathbf{y},\omega)\approx \frac{2
\mbox{i}
e^{[\mbox{i}k_1^iy_1+\mbox{i}k_2^{0,i}\left(y_2-2a\right)]}\alpha^{1,i}\alpha^{0,i}
} { k_2^{1,i}\left(2\alpha^{0,i}\alpha^{1,i}\cos\left(k_2^{1,i} l
\right)-\mbox{i} \left((\alpha^{0,i})^2+(\alpha^{1,i})^2
\right)\sin\left(k_2^{1,i} l \right) \right)^{2}}  \times
\\
\displaystyle \left[\frac{(\alpha^{1,i})^2-(\alpha^{0,i})^2}{2}(k^1)^2 \int_{b}^{a} \chi(x_2)dx_2+
\frac{(\alpha^{1,i})^2+(\alpha^{0,i})^2}{2}(k^1)^2 \int_{b}^{a} \chi(x_2)\cos\left(2k_2^{1,i}
\left(x_2-b\right) \right)dx_2+\right.\\[10pt]
\displaystyle\mbox{i}\alpha^{1,i}\alpha^{0,i} k_{2}^{1,i}\int_{b}^{a}\frac{1}{\rho(x_2)}
\frac{\partial \rho(x_2)}{\partial x_2}\cos\left(2k_2^{1,i}\left(x_2-b\right) \right)dx_2-
\\
\mbox{i}\alpha^{1,i}\alpha^{0,i}(k^1)^2\int_{b}^{a}\chi(x_2)\sin\left(2k_2^{1,i}\left(x_2-b\right)
\right)dx_2+
\\
\left.
\frac{(\alpha^{1,i})^2+(\alpha^{0,i})^2}{2}k_{2}^{1,i}\int_{b}^{a}\frac{1}
{\rho(x_2)}\frac{\partial \rho(x_2)}{\partial
x_2}\sin\left(2k_2^{1,i}\left(x_2-b\right) \right)dx_2 \right] ~,
\label{section311e5}
\end{multline}
where $\displaystyle
\chi(x_2)=\left(\frac{(k(x_2)^2)}{(k^{1})^2}-1 \right)$ is the
contrast function.

We define the average value, cosine transform and  sinus transform
of a function $f(x_2)=h(x_2) \Pi(b\leq x_2 \leq a)$ (wherein
$\Pi(b\leq x_2 \leq a)$ is the so-called gate function and
$l=a-b$) by:
\begin{equation}
\begin{array}{l}
\displaystyle <f(x_2)>=<h(x_2)>=\int_{b}^{a}h(x_2)\frac{dx_2}{l}\\[10pt]
\displaystyle \mbox{TF}_{\mbox{c}}(f(x_2),q)=\int_{-\infty}^{\infty} f(x_2)\cos(q x_2)dx_2\\[10pt]
\displaystyle \mbox{TF}_{\mbox{s}}(f(x_2),q)=\int_{-\infty}^{\infty} f(x_2)\sin(q x_2)dx_2
\end{array}
\label{section311e6}
\end{equation}
respectively.

Eq. (\ref{section311e5}) can then be written in the form:
\begin{equation}
\begin{array}{l}
\displaystyle
p(\mathbf{y},\omega)-p^{0+}(\mathbf{y},\omega)\approx \frac{2
\mbox{i}
e^{[\mbox{i}k_1^iy_1+\mbox{i}k_2^{0,i}\left(y_2-2a\right)]}\alpha^{1,i}\alpha^{0,i}
} { k_2^{1,i}\left(2\alpha^{0,i}\alpha^{1,i}\cos\left(k_2^{1,i} l
\right)-\mbox{i} \left((\alpha^{0,i})^2+(\alpha^{1,i})^2
\right)\sin\left(k_2^{1,i} l \right) \right)^{2}} \times\\[10pt]
\displaystyle
\left[\frac{(\alpha^{1,i})^2-(\alpha^{0,i})^2}{2}(k^1)^2 l
<\chi(x_2)>+\right.
\\[10pt]
\displaystyle\frac{(\alpha^{1,i})^2+(\alpha^{0,i})^2}{2}(k^1)^2 \mbox{TF}_{\mbox{c}}
\left(\chi(x_2-b),2k_2^{1,i}\right)+\mbox{i}\alpha^{1,i}\alpha^{0,i} k_{2}^{1,i}  \mbox{TF}_
{\mbox{c}}\left(\frac{1}{\rho}\frac{\partial \rho}{\partial x_2}(x_2-b),2 k_2^{1,i} \right)-
\\[10pt]
\displaystyle \left. \mbox{i}\alpha^{1,i}\alpha^{0,i}(k^1)^2
\mbox{TF}_{\mbox{s}} \left(\chi(x_2-b),2k_2^{1,i}\right)
+\frac{(\alpha^{1,i})^2+(\alpha^{0,i})^2}{2}k_{2}^{1,i}
\mbox{TF}_{\mbox{s}}\left(\frac{1}{\rho}\frac{\partial \rho}
{\partial x_2}(x_2-b),2 k_2^{1,i} \right)   \right] ~.
\end{array}
\label{section311e7}
\end{equation}

The first-order Born approximation of the reflected field in the
SGF formulation involves the average value of $\chi(x_2)$ and both
the cosine and sinus transform of $\chi(x_2-b)$ and of
$\displaystyle \frac{1}{\rho(x_2-b)}\frac{\partial
\rho(x_2-b)}{\partial x_2}$, while the first order Born
approximation in the FSGF formulation involves only the Fourier
transform of $\chi(x_2-b)$ and of $\displaystyle
\frac{1}{\rho(x_2-b)}\frac{\partial \rho(x_2-b)}{\partial x_2}$
defined by reference to  the material parameter of the host
\cite{innanen}.
%fin d'ajout
%%%%%%%%%%%%%%%%%%%%%%%%%%%%%%%%%%%%%%%%%%%%%%%%%%%%%%%%%%%%%%%%%%%%%%
\subsubsection{First-order Born approximation in the SGF formulation
for $\mathbf{y}\in \Omega_{1}$.}\label{section312}
When $\mathbf{y}\in \Omega_{1}$,
$g_{1}(\mathbf{y},\mathbf{x},\omega)=g_{1}^{1}(\mathbf{y},\mathbf{x},\omega)$,
so that, proceding as previously, we get
\begin{multline}
p(\mathbf{y},\omega)-p^{1}(\mathbf{y},\omega)
\approx\int_{-\infty}^{+\infty} \int_{b}^{a}
g_{1}^{1}(\mathbf{y},\mathbf{x})\left[\left(k(x_2)^2-(k^1)^2\right)-\frac{1}
{\rho(x_2)}\frac{\partial \rho(x_2)}{\partial x_2}\,\frac{\partial
}{\partial x_2}  \right] p^{1}(\mathbf{x})\,dx_1\, dx_2
\\
\approx\int_{b}^{a} \Big[ \Big\{\frac{\mbox{i}}{2 k_2^{1,i}}
e^{[\mbox{i}k_1^i y_1+\mbox{i}k_2^{1,i}|y_2-x_2|]}
+\mbox{i}e^{\mbox{i}k_1^iy_1}\times
\\
\frac{\left(\left((\alpha^{1,i})^2-
(\alpha^{0,i})^2\right)\cos\left(k_2^{1,i}\left(y_2+x_2-a-b\right)
\right)+
e^{\mbox{i}k_{2}^{1,i}l}\left(\alpha^{0,i}-\alpha^{1,i}\right)^2\cos\left(k_2^{1,i}\left(y_2-x_2\right)
\right) \right) }{2 k_2^{1,i}
\left(2\alpha^{0,i}\alpha^{1,i}\cos\left(k_2^{1,i} l
\right)-\mbox{i}\left((\alpha^{0,i})^2+(\alpha^{1,i})^2
\right)\sin\left(k_2^{1,i} l \right) \right)}\Big\}
 \Big] \times
 \\
\left[\left(k(x_2)^2-(k^1)^2\right)-\frac{1}{\rho(x_2)}\frac{\partial
\rho(x_2)}{\partial x_2}\,\frac{\partial }{\partial x_2}
\right]\tilde{p}^{1}(x_2) \, dx_2 ~.
\label{section312e1}
\end{multline}
%%%%%%%%%%%%%%%%%%%%%%%%%%%%%%%%%%%%%%%%%%%%%%%%%%%%%%%%%%%%%%%%%%%%%%
\subsubsection{First-order Born approximation in the SGF formulation
for $\mathbf{y}\in \Omega_{0-}$.}\label{section313}
When $\mathbf{y}\in \Omega_{0-}$,
$g_{1}(\mathbf{y},\mathbf{x},\omega)=g_{1}^{0^-}(\mathbf{y},\mathbf{x},\omega)$,
so that
\begin{equation}
\begin{array}{l}
\displaystyle p(\mathbf{y},\omega)-p^{0-}(\mathbf{y},\omega)
\approx\int_{-\infty}^{+\infty}\int_{b}^{a}
g_{1}^{0-}(\mathbf{y},\mathbf{x})\left[\left(k(x_2)^2-(k^1)^2\right)-
\frac{1}{\rho(x_2)}\frac{\partial
\rho(x_2)}{\partial x_2}\,\frac{\partial }{\partial x_2}
\right]p^{1}(\mathbf{x})\,dx_1\, dx_2\\[10pt]
\displaystyle \approx\int_{b}^{a}
\left[\frac{\mbox{i}e^{[\mbox{i}k_1^iy_1+\mbox{i}k_2^{0,i}\left(b-y_2\right)]}
\alpha^{1,i}\left(\alpha^{1,i}\cos\left(k_2^{1,i}\left(a-x_2\right)
\right)-\mbox{i}\alpha^{0,i}\sin\left(k_2^{1,i}\left(a-x_2\right)
\right) \right) }{ k_2^{1,i}
\left(2\alpha^{0,i}\alpha^{1,i}\cos\left(k_2^{1,i} l
\right)-\mbox{i}\left((\alpha^{0,i})^2+(\alpha^{1,i})^2
\right)\sin\left(k_2^{1,i} l \right) \right)}\right]
\times
\\
\displaystyle \left[\left(k(x_2)^2-(k^1)^2\right)-\frac{1}{\rho(x_2)}\frac{\partial
\rho(x_2)}{\partial x_2}\,\frac{\partial }{\partial x_2}
\right]\tilde{p}^{1}(\mathbf{x})\, dx_2 ~.
\end{array}
\label{section313e1}
\end{equation}

Introducing the expression of $\tilde{p}^{1}(\mathbf{x})$ from (\ref{appendix1e2}), and by making use of the definition (\ref{section311e6}), the previous equation can be written in the form:
\begin{multline}
p(\mathbf{y},\omega)-p^{0-}(\mathbf{y},\omega) \approx
\left[\frac{2
\mbox{i}e^{[\mbox{i}k_1^iy_1-\mbox{i}k_2^{0,i}\left(l+y_2\right)]}
\alpha^{1,i}\alpha^{0,i} }{ k_2^{1,i}
\left(2\alpha^{0,i}\alpha^{1,i}\cos\left(k_2^{1,i} l
\right)-\mbox{i}\left((\alpha^{0,i})^2+(\alpha^{1,i})^2
\right)\sin\left(k_2^{1,i} l \right) \right)^2}\right] \times
\\
\Big[
\left(\frac{(\alpha^{1,i})^2+(\alpha^{0,i})^2}{2}\cos\left(k_{2}^{1,i}l
\right)-\mbox{i}\alpha^{0,i}\alpha^{1,i}\cos\left(k_{2}^{1,i}l
\right) \right) l (k^1)^2 < \chi(x_2)>
\\
\left(\mbox{i}\alpha^{0,i}\alpha^{1,i}\cos\left(k_{2}^{1,i}l
\right)+
\frac{(\alpha^{1,i})^2+(\alpha^{0,i})^2}{2}\cos\left(k_{2}^{1,i}l
\right) \right)k_{2}^{1,i}l < \frac{1}{\rho(x_2)} \frac{\partial
\rho(x_2)}{\partial x_2}> +
\\
\frac{(\alpha^{1,i})^2-(\alpha^{0,i})^2}{2} \left(
\cos\left(k_2^{1,i}l\right)(k^1)^2
\mbox{TF}_{\mbox{c}}\left(\chi(x_2-b),2k_2^{1,i}\right)+\sin\left(k_2^{1,i}l\right)(k^1)^2
\mbox{TF}_{\mbox{s}}\left(\chi(x_2-b),2k_2^{1,i}\right)\right)
\\
\frac{(\alpha^{1,i})^2-(\alpha^{0,i})^2}{2}
\Big(
\cos\left(k_2^{1,i}l\right)k_2^{1,i}\mbox{TF}_{\mbox{s}}\left(\frac{1}{\rho}\frac{\partial
\rho}{\partial x_2}(x_2-b),2 k_2^{1,i} \right)-
\\
\sin\left(k_2^{1,i}l\right)
k_2^{1,i}\mbox{TF}_{\mbox{c}}\left(\frac{1}{\rho}\frac{\partial
\rho}{\partial x_2}(x_2-b),2 k_2^{1,i} \right) \Big)
\Big] ~.
\label{section313e5}
\end{multline}
This equation involves the average values, the cosine and sinus
transform of both $\chi(x_2)$ and $\displaystyle
\frac{1}{\rho}\frac{\partial \rho}{\partial x_2}$. Compared with
the formulae (\ref{section311e7}), the transmitted field involves
the additional term corresponding to the average value of
$\displaystyle \frac{1}{\rho}\frac{\partial \rho}{\partial x_2}$.
%fin d'ajout
%%%%%%%%%%%%%%%%%%%%%%%%%%%%%%%%%%%%%%%%%%%%%%%%%%%%%%%%%%%%%%%%%%%
\subsection{The iterative scheme for solving the direct problem}\label{section32}
As pointed out previously, our aim is to define an iterative
scheme to solve the direct problem of the diffraction of an
incident plane wave by a heterogeneous porous slab. This would be
of great interest for both
 the direct and inverse problems, due to the possible increased accuracy it can enable with
 respect to both the zeroth- and first-order Born approximations (in both the SGF and FSGF formulations).

We want to compute the total  fields in $\Omega_{0^+}$ and
$\Omega_{0^-}$. These problems being formally similar, we will
only  detail the computation  of
$p(\mathbf{y})~;~\mathbf{y}\in\Omega_{0^+}$.

Let  $p^{0^{+}(j)}(\mathbf{y})$ and $p^{1(j)}(\mathbf{y})$
designate the $j$-th iterates  of the pressure fields in
$\Omega_{0^+}$ and $\Omega_{1}$ respectively. The iterative scheme
proceeds as follows:
\begin{itemize}
\item Calculation of $p^{0^{+}(1)}(\mathbf{y})$ through (\ref{section311e3}),
corresponding to the application of the Born approximation in the
SGF formulation.
\item Calculation of $p^{0^{+}(j)}(\mathbf{y})$ for $j>1$.
\end{itemize}
More specifically, we first have to calculate the pressure field
in $\Omega_1$ by means of
\begin{equation}
\hspace{-1.0cm}\begin{array}{l}
\displaystyle p^{1(j)}(\mathbf{y},\omega)-p^{1(0)}(\mathbf{y},\omega) \approx\int_{b}^{a} \left[
\left\{\frac{\mbox{i}}{2 k_2^{1,i}} e^{\mbox{i}k_1^i y_1+\mbox{i}k_2^{1,i}|y_2-x_2|}+  \right.\right.
\\[10pt]
\displaystyle \left.\left. \frac{\mbox{i}e^{\mbox{i}k_1^iy_1}\left(\left((\alpha^{1,i})^2-
(\alpha^{0,i})^2\right)\cos\left(k_2^{1,i}\left(y_2+x_2-a-b\right) \right)+e^{\mbox{i}k_{2}^{1,i}l}
\left(\alpha^{0,i}-\alpha^{1,i}\right)^2\cos\left(k_2^{1,i}\left(y_2-x_2\right) \right) \right) }
{2 k_2^{1,i} \left(2\alpha^{0,i}\alpha^{1,i}\cos\left(k_2^{1,i} l \right)-\mbox{i}\left((\alpha^{0,i})^2+
(\alpha^{1,i})^2 \right)\sin\left(k_2^{1,i} l \right) \right)}\right\}
 \right] \times\\[10pt]
\displaystyle \left[\left(k(x_2)^2-(k^1)^2\right)-\frac{1}{\rho(x_2)}\frac{\partial \rho(x_2)}
{\partial x_2}\,\frac{\partial }{\partial x_2}  \right]\tilde{p}^{1(j-1)}(x_2) \, dx_2
\end{array}
\label{section32e1}
\end{equation}
wherein $p^{1(0)}(\mathbf{y},\omega)$ is the expression given in
(\ref{appendix1e2}).

Once a new $p^{1(j)}(\mathbf{y},\omega)$ is evaluated, one
computes a new $p^{0^{+}(j)}(\mathbf{y})$ by means of the
relation
\begin{multline}
p^{0^+(j)}(\mathbf{y},\omega)-p^{0+(0)}(\mathbf{y},\omega) \approx
\\
\int_{b}^{a} \left[\frac{\mbox{i}
e^{\mbox{i}k_1^iy_1+\mbox{i}k_2^{0,i}\left(y_2-a\right)}
\alpha^{1,i}\left(\alpha^{1,i}\cos\left(k_2^{1,i}\left(x_2-b\right)
\right)-\mbox{i}\alpha^0 \sin\left(k_2^{1,i}\left(x_2-b\right)
\right) \right) }{ k_2^{1,i} \left(2\alpha^{0,i}
\alpha^{1,i}\cos\left(k_2^{1,i} l
\right)-\mbox{i}\left((\alpha^{0,i})^2+(\alpha^{1,i})^2 \right)
\sin\left(k_2^{1,i} l \right) \right)} \right] \times\\[10pt]
\displaystyle
\left[\left(k(x_2)^2-(k^1)^2\right)-\frac{1}{\rho(x_2)}\frac{\partial
\rho(x_2)} {\partial x_2}\,\frac{\partial }{\partial x_2}
\right]\tilde{p}^{1(j)}(x_2) \, dx_2 ~.
\label{section32e2}
\end{multline}
{\bf{Remark:}} Another scheme is  the iterative calculation of
$p^{1(j)}(\mathbf{y},\omega)$ and the subsequent computation of
the reflected field $p^{0^+(j)}(\mathbf{y},\omega)$.
\newline
\newline
The differentiation of $\tilde{p}^{1(j)}$, which is a particular
feature of our method, is carried out analytically for $j=0$ by
means of (\ref{appendix1e2b}), and numerically for $j\geq 1$ using
the finite difference scheme:
\begin{equation}
\frac{\partial}{\partial x_2}\tilde{p}^{1(j)}(x_2)\approx
\frac{\tilde{p}^{1(j)}(i+1)-
\tilde{p}^{1(j)}(i)}{X_2(i+1)-X_2(i)} \label{section32e3}
\end{equation}
The computation  of $\displaystyle \left. \frac{\partial
}{\partial x_2}\tilde{p}^{1(j)}(x_2)\right|_{x_2=X_2(N)=a}$ cannot
be carried out in this manner. We approximate this derivative by
using the fact that $\displaystyle \frac{1}{\rho(x_2)}\frac{\partial
\tilde{p}^{1(j)}(x_2)}{\partial x_2}$ is conserved, so that
\begin{equation}
\left. \frac{\partial }{\partial x_2}\tilde{p}^{1(j)}(x_2)\right|_{x_2=x_2(N)=a} \approx
\frac{\rho(N)}{\rho(N-1)}\left. \frac{\partial }{\partial x_2}
\tilde{p}^{1(j)}(x_2)\right|_{x_2=X_2(N-1)}
\label{section32e4}
\end{equation}
wherein $N$ is the number of discretisation points used to
performed the calculation.
%%%%%%%%%%%%%%%%%%%%%%%%%%%%%%%%%%%%%%%%%%%%%%%%%%%%%%%%%%%%%%%%%%%%%
\section{Outline of the numerical procedure}\label{section4}
We focus on the response of a double layer (each layer being
homogeneous) porous slab (called layer1 and layer2), considered to
be a single inhomogeneous slab. We assume that the medium in the
slab responds to a solicitation as does  an equivalent fluid
(i.e., this is the rigid-frame approximation).

In an equivalent fluid medium, \cite{Deryck} the appropriate
conservation of momentum and constitutive relations take the form:
\begin{equation}
\omega^{2} p +\frac{1}{\kappa_{e}(\mathbf{x},\omega)}\nabla \cdot \left(\frac{1}
{\rho_{e}(\mathbf{x},\omega)}\nabla p \right)=0
\label{section4e1}
\end{equation}
wherein
\begin{equation}
\begin{array}{l}
\displaystyle \rho_{e}(\mathbf{x},\omega)=\rho_{e}(x_2,\omega)=
\frac{\rho_f \alpha_{\infty}(x_2)}{\phi(x_2)}\left(1+\mbox{i}
\frac{\omega_{c}(x_2)}{\omega}F(x_2,\omega) \right)
\\[12pt]
\displaystyle \frac{1}{\kappa_{e}(\mathbf{x},\omega)}=\frac{1}{\kappa_{e}(x_2,\omega)}=
\frac{\gamma P_{0}}{\phi(x_2)\left(\gamma-(\gamma-1)
\left(1+\mbox{i}\frac{\omega_{c}(x_2)}{\mbox{Pr}^2 \omega}G(x_2,\mbox{Pr}^2 \omega)
\right)^{-1}\right)}
\end{array}
\label{section4e2}
\end{equation}
with $\displaystyle w_c(x_2)=\frac{\sigma(x_2)\phi(x_2)}{\rho_f
\alpha(x_2)}$ and $G(x_2,\mbox{Pr}^2 \omega)$
\cite{AllardChampoux}, $F(x_2,\omega)$\cite{johnson} being two
relaxation functions given by
\begin{equation}
\begin{array}{l}
\displaystyle F(x_2,\omega)=\sqrt{1-\mbox{i}\frac{4 \eta \rho_{f}
\alpha_{\infty}(x_2)^2}{\sigma(x_2)^2\phi(x_2)^2
\Lambda(x_2)^2}\omega}\\[8pt]
\displaystyle G(x_2,\mbox{Pr}^2\omega)=\sqrt{1-\mbox{i}\frac{4 \eta \rho_{f}
\alpha_{\infty}(x_2)^2}{\sigma(x_2)^2\phi(x_2)^2\Lambda'(x_2)^2}\mbox{Pr}^2\omega}
\end{array}
\label{section4e3}
\end{equation}
The chosen profile of porosity $\phi(x_2)$,
$\Lambda(x_2)$, $\Lambda'(x_2)$,
$\alpha(x_2)$ and $\sigma(x_2)$ is presented table \ref{tab1}.

\begin{table}[h]
\centering\begin{tabular}{lccccccc}
\hhline{========}
 &  & $\phi$ & $\tau_{\infty}$ & $\Lambda$ & $\Lambda'$ & $R_f$ & Thickness \\
 &        &                   & $(\mu m)$ & $(\mu m) $&$(Ns.m^{-4})$& $(mm)$ \\
\hline
 \textbf{Layer 1} &  & 0.96     & 1.07  & 273 & 672 & 2843  & 7.1  \\
 \textbf{Layer 2} &  & 0.99 & 1.001 & 230 & 250 & 12000 & 10.0\\
\hhline{========}
\end{tabular}
\label{tab1}
\caption{\emph{Properties of the two-layer medium studied.}}
\end{table}

The inhomogeneous porous slab is included between $b=-10\times
10^{-3}m$ and $a=7.1 \times 10^{-3}m$. The  contact surface
between the two homogeneous porous sub-slabs, is located at
$x_2=0m$.

Special attention must be paid to:
\begin{itemize}
\item  the discretisation of $x_{2}$ in order to correctly model eventual
jumps,
\item the modeling of the jump; as pointed out in appendix \ref{appendixe2},
the spatial dependence of the density $\rho(\mathbf{x})$ can lead
to meaningless integrals, especially when this parameter presents
some discontinuities. To avoid this problem, we will consider such
jumps to be well-approximated by the continuous function
\begin{equation}
H(x_2-e)\approx \frac{1}{2}\left(1+\mbox{erf}\left(\frac{x_2-e}{s} \right) \right)
\label{section4e4}
\end{equation}
where $e$ is the location of the jump, $s$  the slope of the
smooth jump and erf the error function,
\item  the determination of the parameters $\rho^{1}(\omega)$ and $k^{1}(\omega)$ filling the initial
homogeneous slab.
\end{itemize}
%%%%%%%%%%%%%%%%%%%%%%%%%%%%%%%%%%%%%%%%%%%%%%%%%%%%%%%%%%%%%%%%%%%%%
\subsection{Choice of the discretization step.}\label{section41}
Because of the necessary correct modeling of the continuous steps,
making use of the formulae (\ref{section4e4}) at both the location
of the step and the sides of the slab, we  consider a logarithmic
scale, with an increase of the point density at these locations.
This logarithmic scale  occurs over a width $\Delta$ on both sides
of a jump. In our computations, $\Delta$ is chosen equal to
$8\times 10^{-6}$.
%%%%%%%%%%%%%%%%%%%%%%%%%%%%%%%%%%%%%%%%%%%%%%%%%%%%%%%%%%%%%%%%%%%%%
\subsection{Choice of the modeling of the jumps}\label{section42}
To model the jumps at $x_2=0$, we define a function $\zeta$ such that :
\begin{equation}
\zeta(x_2)=\zeta_1+\frac{(\zeta_2-\zeta_1)}{2}(1+\mbox{erf}(\frac{x_2-e}{s}))
\label{section42e1}
\end{equation}
where $\zeta(x_2)$ can  be $\phi(x_2)$, $\lambda(x_2)$,
$\lambda'(x_2)$, $\alpha(x_2)$, or $\sigma(x_2)$, and the indices
$_1$ and $_2$ refer to the values of the parameter $\zeta$ of the
homogeneous layer 1 or 2. The quantities $\rho(x_2)$ and $k(x_2)$
are then computed.

Once $\rho^1$ and $k^1$ are determined, in order to take into
account  the jumps at both (or at least one) sides of the entire
slab, we compute:
\begin{equation}
\rho(x_2)=\rho^1+(\rho(x_2)-\rho^1)\left(\mbox{erf}(\frac{x_2-b}{s})-\mbox{erf}(\frac{x_2-a}{s})-1\right)
~. \label{section42e2}
\end{equation}
Thus, on both sides of the slab we model the ``half'' jump from
$\rho^1$ to $\rho(x_2)$ using a half of the erf function (compared
to the jump inside the entire slab). This constitutes a better fit
of the real jump.

In all our computations, the parameter $s$ is chosen equal to
$2\times 10^{-6}$.
%%%%%%%%%%%%%%%%%%%%%%%%%%%%%%%%%%%%%%%%%%%%%%%%%%%%%%%%%%%%%%%%%%%%%
\subsection{Choice of parameters $\rho^{1}(\omega)$ and $k^{1}(\omega)$.}\label{section43}
The purpose of the SFG is to reduce the kernel of the integral
(\ref{section2e4}) compared to the kernel of the integral
(\ref{section2e5}) in the FSGF formulation. Because of the spatial
dependence of the density, the integral (\ref{section2e4}) can be
split into two integrals whose respective kernels are:
\begin{equation}
\displaystyle \left(k(\mathbf{x})^{2}-(k^1)^2\right)\mbox{ and }
\frac{1}{\rho}\frac{\partial \rho}{\partial x_2} ~.
\end{equation}
The easiest way to reduce these kernels would be (referred-to as
choice 1.), all the characteristic parameters of the slab being
known, to consider the average value of $\rho(x_2,\omega)$ and
$k(x_2,\omega)$ over $x_2\in[b,a]$, as shown figure \ref{section43f1}
for $\Re(\rho(x_2),175kHz)$.

Another choice (referred-to as choice 2.), which can be more
convenient,  consists in taking $\rho^{1}(\omega)$ and
$k^{1}(\omega)$ equal to the minimal value of $\rho(x_2,\omega)$
and $k(x_2,\omega)$ over $x_2\in[b,a]$, as shown
figure \ref{section43f1b}. This choice would normally lead to the
disappearance of the remanent density (i.e. equal to $\rho^{1}$
whose values is larger than $\rho(b)$) at $x_2=b$,
figure \ref{section43f1}. Another advantage of this choice is the
reduction of the interval of  integration , the first kernel
vanishing over a part of this interval.

%\newline
\begin{minipage}{7.5cm}
\begin{figure}[H]
\centering\psfig{figure=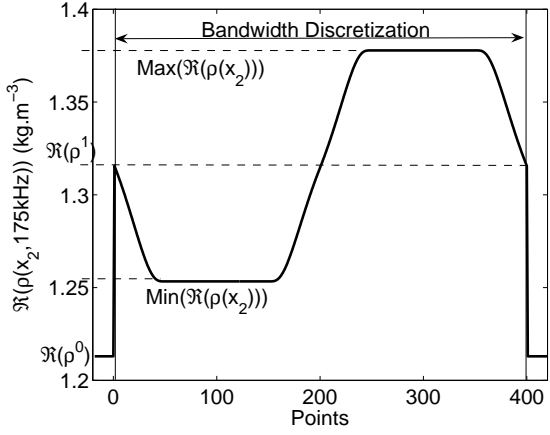,width=8.0cm}
\caption{Real part of the density profile corresponding to $\rho^{1}(175kHz)$ chosen as the average value of $\rho(x_2,175kHz)$ over $x_2\in[b,a]$.}
\label{section43f1}
\end{figure}
\end{minipage}\hfill
\begin{minipage}{7.5cm}
\begin{figure}[H]
\centering\psfig{figure=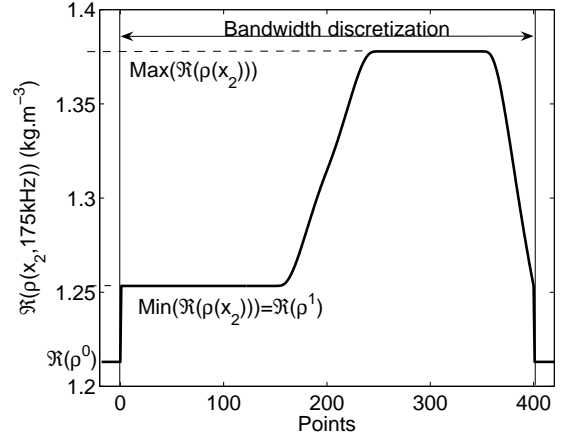,width=8.0cm}
\caption{Real part of the density profile corresponding to $\rho^{1}(175kHz)$ chosen as the minimal value of $\rho(x_2,175kHz)$ over $x_2\in[b,a]$.}
\label{section43f1b}
\end{figure}
\end{minipage}
\newline

\begin{figure}[H]
\begin{minipage}{8.0cm}
\centering{Incidence angle of $\displaystyle 0$}
\end{minipage}\hfill
\begin{minipage}{8.0cm}
\centering{Incidence angle of $\displaystyle \frac{\pi}{3}$}
\end{minipage}\hfill
\begin{minipage}{8.0cm}
\centering\psfig{figure=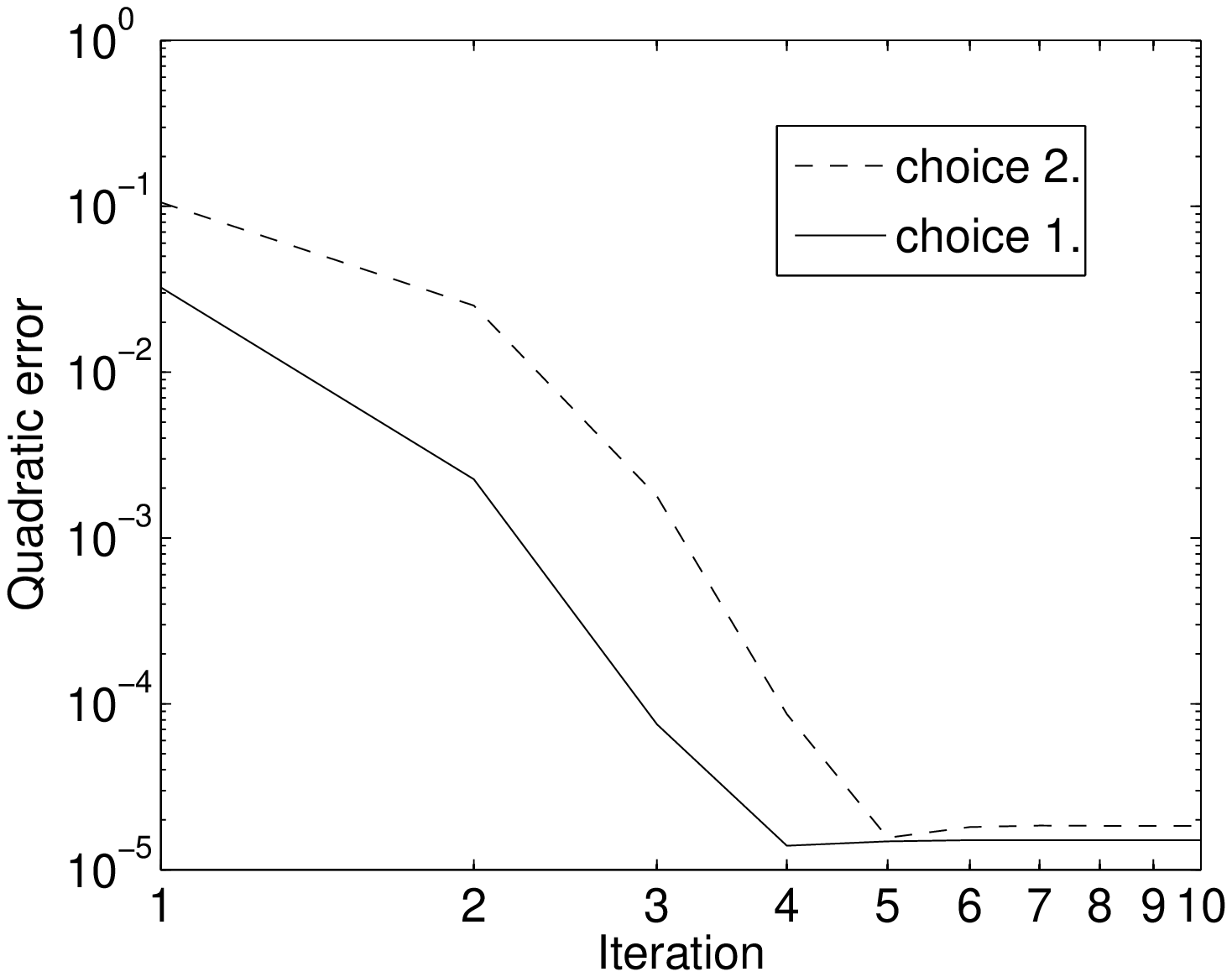,width=8.0cm}
\end{minipage}\hfill
\begin{minipage}{8.0cm}
\centering\psfig{figure=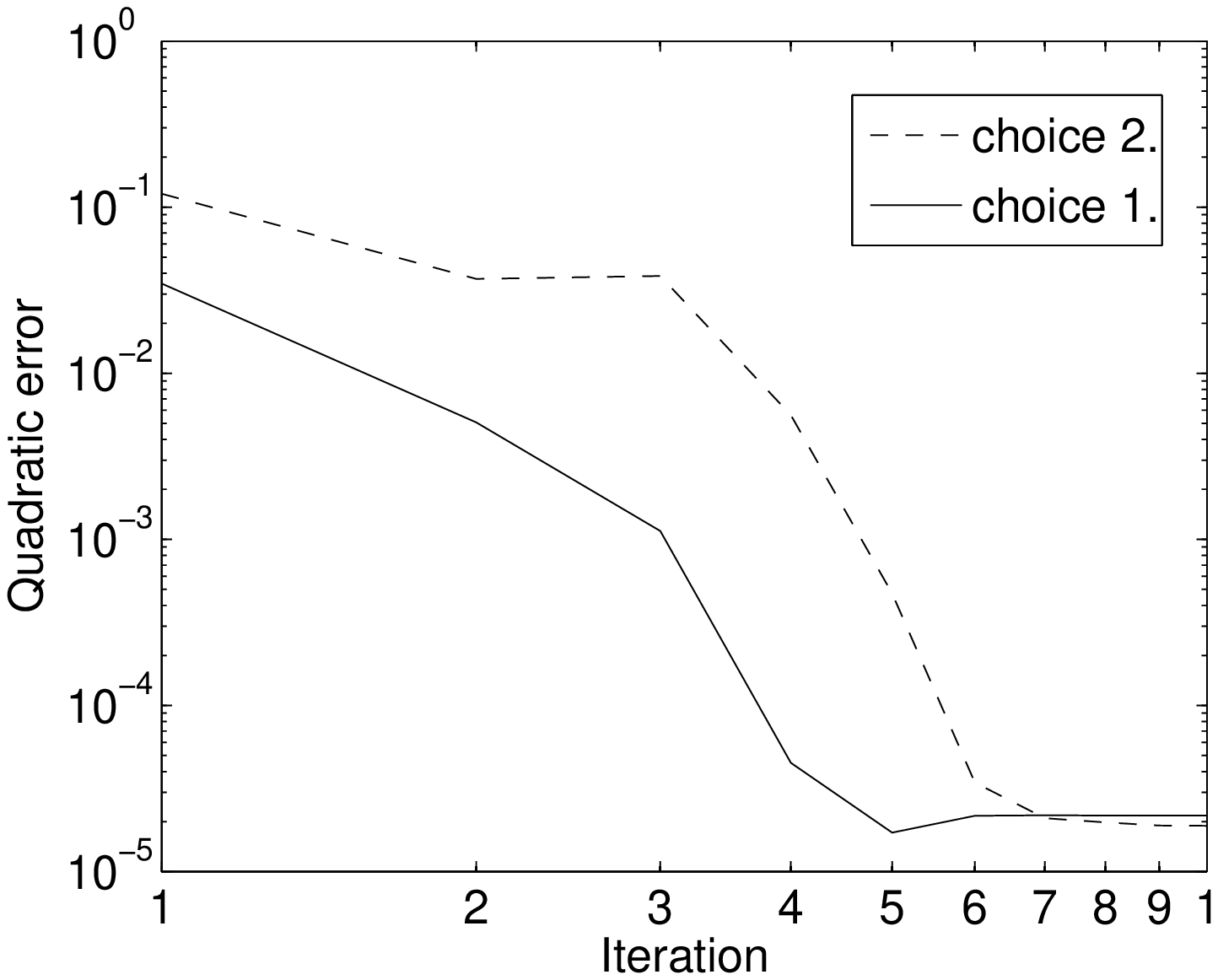,width=8.0cm}
\end{minipage}
\caption{Evolution of the quadratic error as a function of the number of iterations.
On the left: angle of incidence $\displaystyle 0$. On the right: $\displaystyle \frac{\pi}{3}$.}
\label{section43f3}
\end{figure}

To give an idea of the accuracy of the method, we introduce the
following measure of the quadratic error, calculated for the
$j$-th iteration:
\begin{equation}
E^{0^+,j}=\frac{\int_{0}^{T}\left(p_{TMM}^{0^+d}(\mathbf{x},t)-p_{SGIM}^{0^+d,i}
(\mathbf{x},t) \right)^2
dt}{\int_{0}^{T}\left(p_{TMM}^{0^+d}(\mathbf{x},t)\right)^2 dt}
\label{section43e1}
\end{equation}
wherein $p_{SGIM}^{0^+d}(\mathbf{x},t)$ is the reflected pressure
as computed by our Specific Green's Function based Iterative
Scheme (SGIM), and $p_{TMM}^{0^+d}(\mathbf{x},t)$  the reflected
pressure as computed by the classical Transfer Matrix Method
(TMM), appendix \ref{appendixe3}.
The  quadratic error corresponding to our computations is given
figure \ref{section43f3}.

These experiments show that the correct choice of
$\rho^{1}(\omega)$ and $k^{1}(\omega)$ is indeed to consider the
average value of $\rho(x_2,\omega)$ and $k(x_2,\omega)$ over
$x_2\in[b,a]$. This choice leads to a quicker and better
convergence than the one obtained by the choice of
$\rho^{1}(\omega)$ and $k^{1}(\omega)$ as the minimum of
$\rho(x_2,\omega)$ and $k(x_2,\omega)$ over $x_2\in[b,a]$.

\begin{figure}[H]
\begin{minipage}{8.0cm}
\centering\psfig{figure=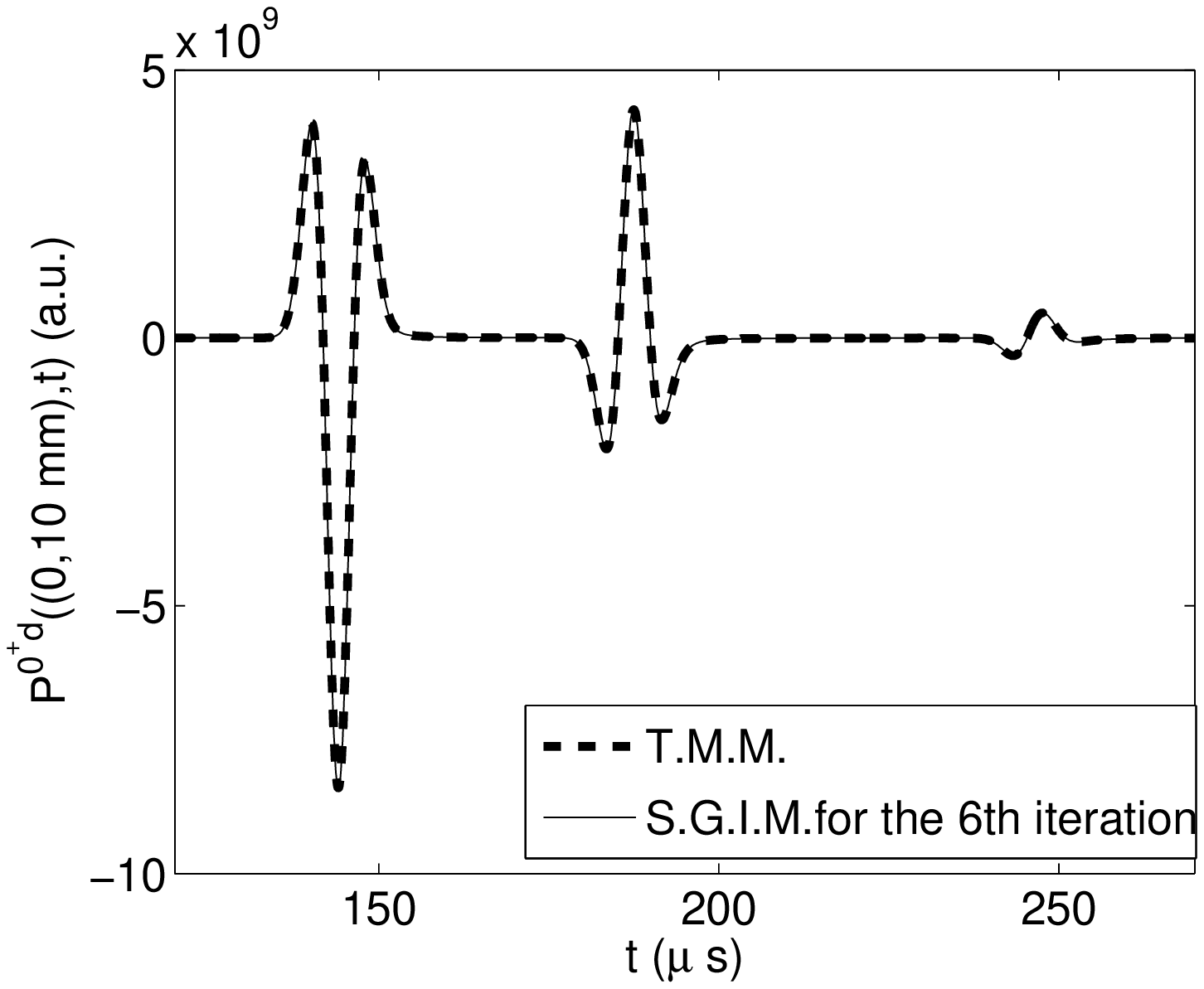,width=7.0cm}
\end{minipage}\hfill
\begin{minipage}{8.0cm}
\centering\psfig{figure=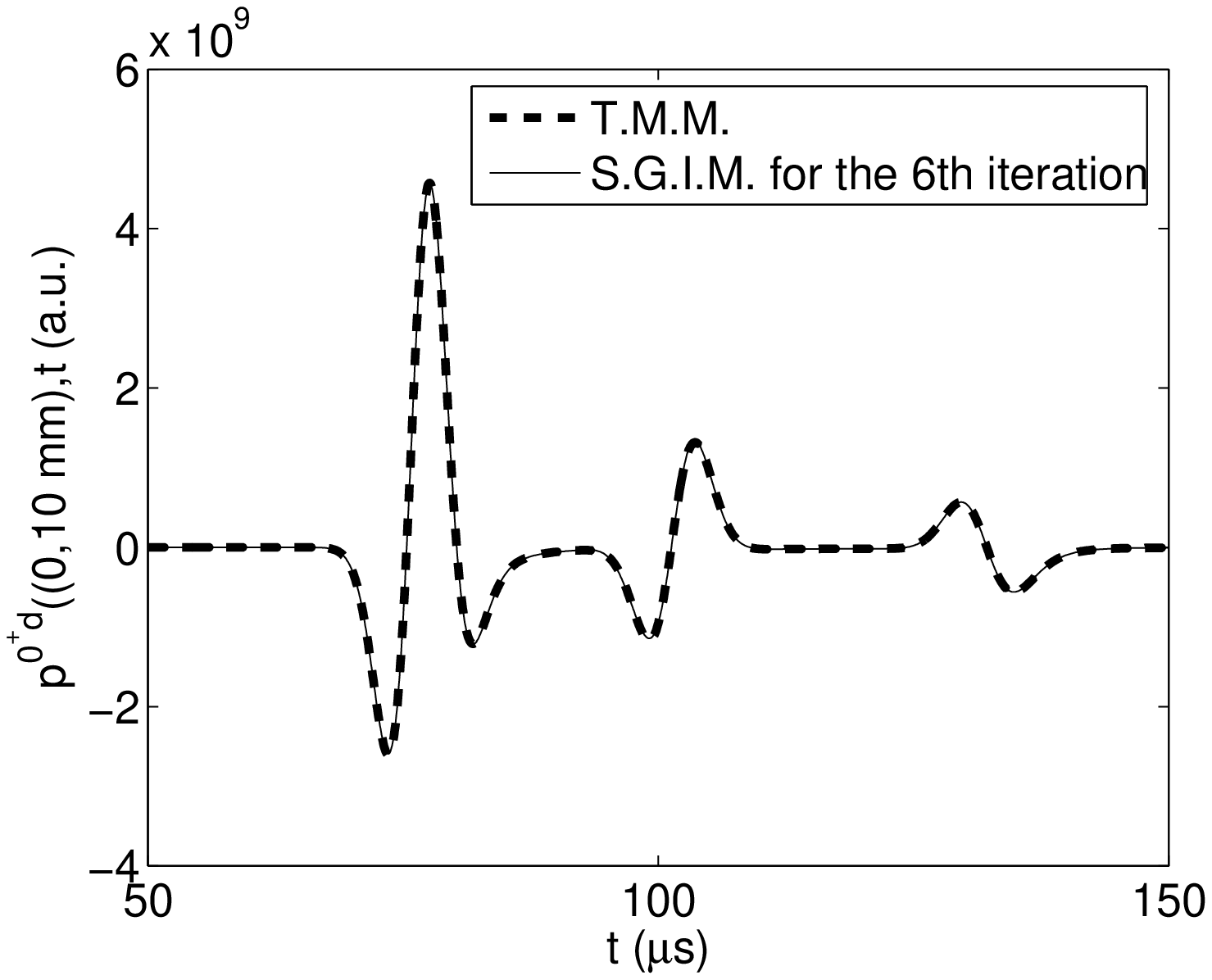,width=7.0cm}
\end{minipage}\hfill
\caption{Reflected pressure as computed by the classical Transfer Matrix Method
(TMM) -dashed curve- and as computed by our Specific Green's Function based Iterative Method (SGIM) when  choice 1. is made. The angle of incidence $\displaystyle 0$ on the left and $\displaystyle \frac{\pi}{3}$ on the right.}
\label{section43f4}
\end{figure}

For both choices of these parameters, after a certain number of
iterations, the SGIM results are  the same as the classical TMM
results, as shown figure \ref{section43f4}, for example, when choice 1
is made.

%Ajout

{\textbf{Remark: }} For both  choices, $\displaystyle
<\frac{1}{\rho}\frac{\partial \rho}{\partial x_2}>$, involved in
the calculation of  the  first order Born approximation of both
the reflected and transmitted fields,  vanishes, i.e.,
\begin{equation}
<\frac{1}{\rho}\frac{\partial \rho}{\partial x_2}> =
\int_{b}^{a}\frac{1}{\rho}\frac{\partial \rho}{\partial x_2} dx_2=
\int_{b}^{a}\frac{\partial \ln\left(\rho (x_2)\right)}{\partial x_2} dx_2=
\ln\left(\frac{\rho (a)}{\rho (b)}\right)=0
\end{equation}
because $\rho (x_2)= \rho^1(\omega)$ for both $x_2=a$ and $x_2=b$.
\newline
\newline
{\textbf{Remark: }} Other choices are possible, such as the one
leading to the disappearance of the averages $<\chi(x_2)>$
involved in the calculation of the first order Born approximation
of both the reflected and transmitted field. Consider
$<\chi(x_2)>$:
\begin{equation}
<\chi(x_2)> =\int_{b}^{a}\left(\frac{(k(x_2))^2}{(k^1)^2}-1 \right)
\frac{dx_2}{l}=\frac{1}{(k^1)^2}\int_{b}^{a}(k(x_2))^2\frac{dx_2}{l}-1
\label{section43e2}
\end{equation}
which vanishes only if $\displaystyle
k^{1}=\sqrt{\int_{b}^{a}(k(x_2))^2\frac{dx_2}{l}}$. This choice
corresponds to the particular case in which $k(x_2)$ is such that
the Schwartz inequality is satisfied:
\begin{equation}
\left(\int_{b}^{a} k(x_2)\frac{dx_2}{l} \right)^2\leq \int_{b}^{a}
(k(x_2))^2 \frac{dx_2}{l}~.
\label{section43e3}
\end{equation}
%%
% Fin ajout
%%%%%%%%%%%%%%%%%%%%%%%%%%%%%%%%%%%%%%%%%%%%%%%%%%%%%%%%%%%%%%%%%%%%%%%%%%%%%%
\section{Results and discussion}\label{section5}
We first present results, as calculated by the iterative scheme
(FGIM) initialized with the zeroth-order Born approximation
arising from the integral formulation incorporating the free-space
Green's function, to emphasize the fact that this method does not
converge in all cases.
\begin{figure}[H]
\begin{minipage}{5.5cm}
\centering\psfig{figure=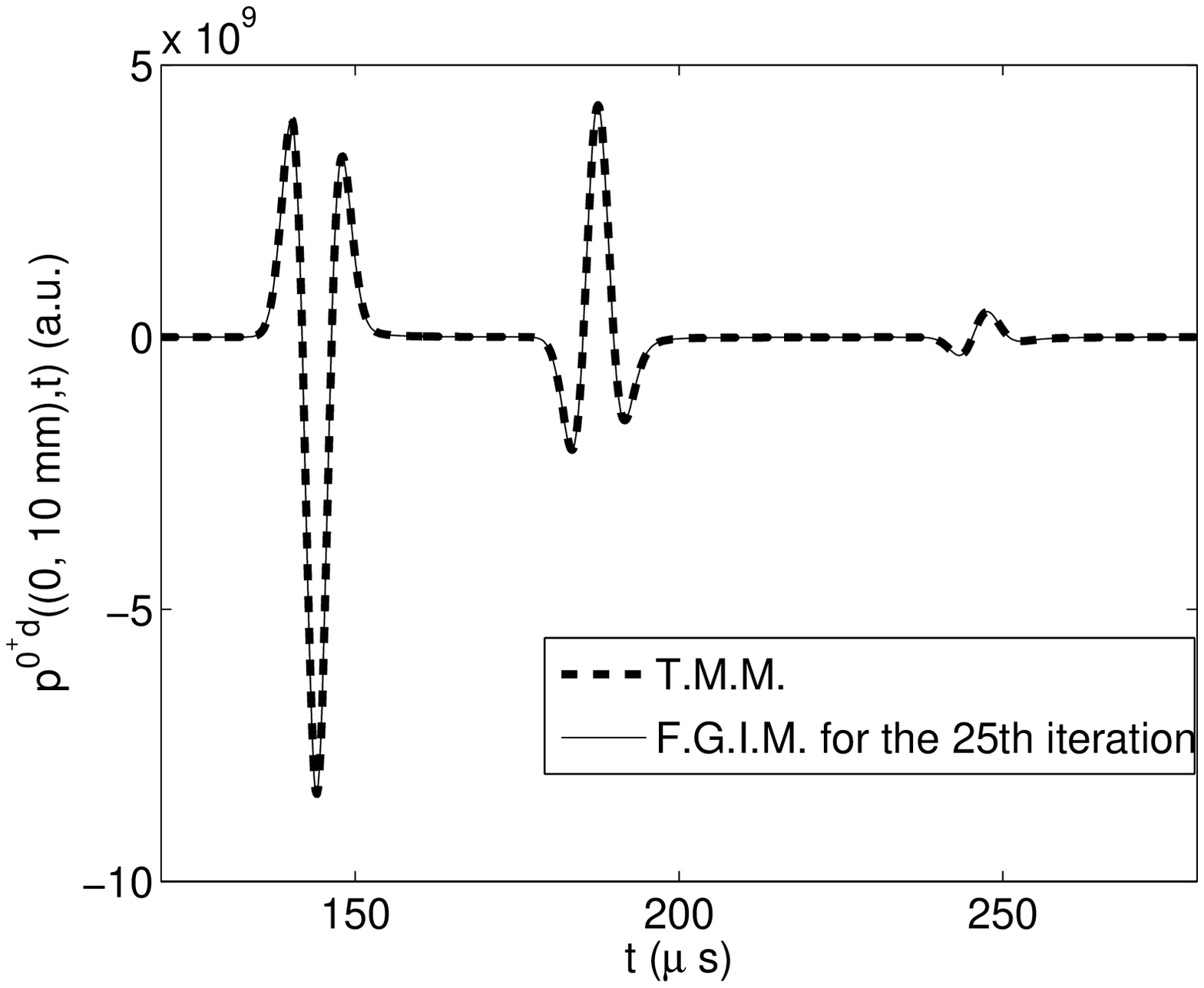,width=5.5cm}
\end{minipage}\hfill
\begin{minipage}{5.5cm}
\centering\psfig{figure=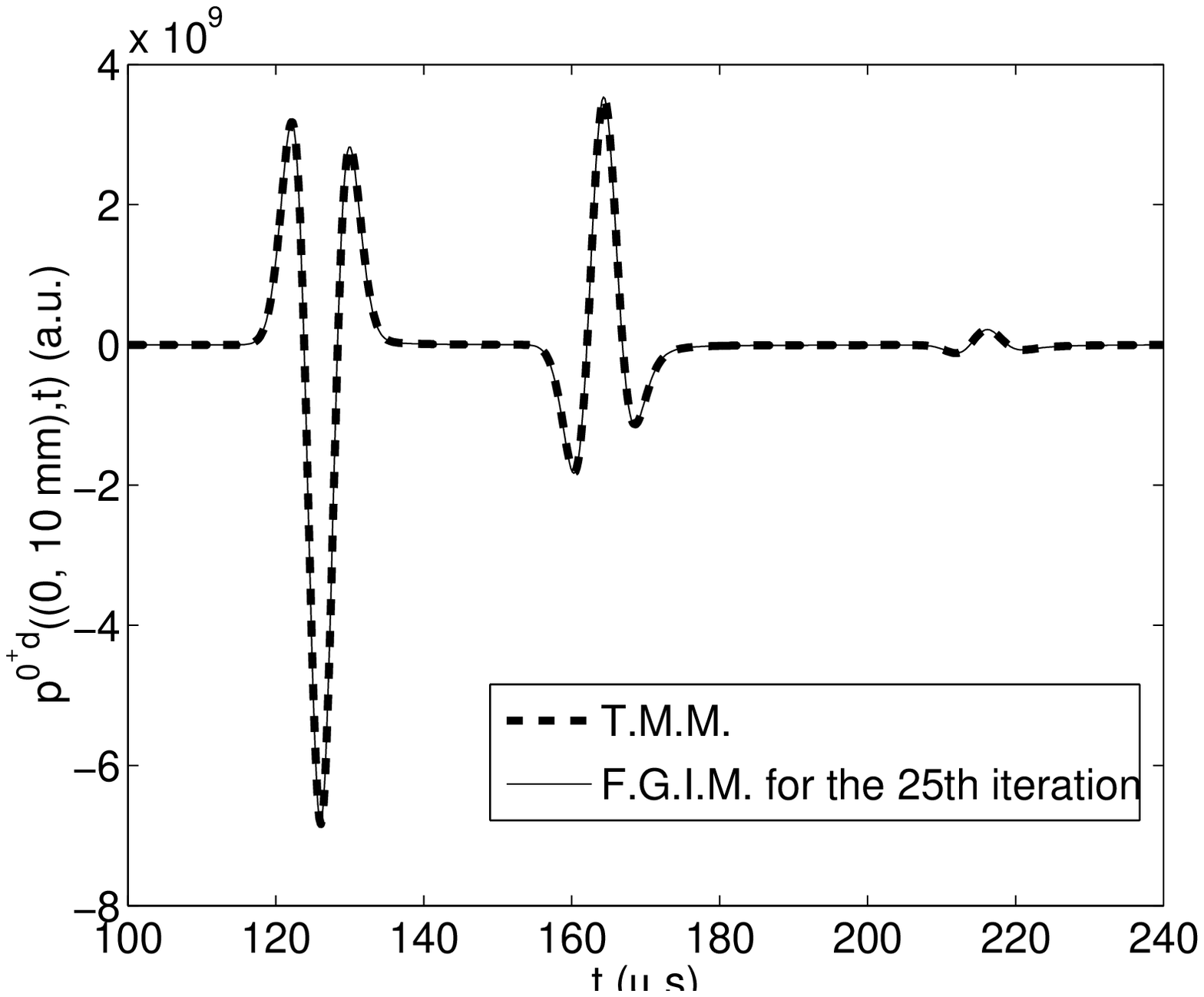,width=5.5cm}
\end{minipage}\hfill
\begin{minipage}{5.5cm}
\centering\psfig{figure=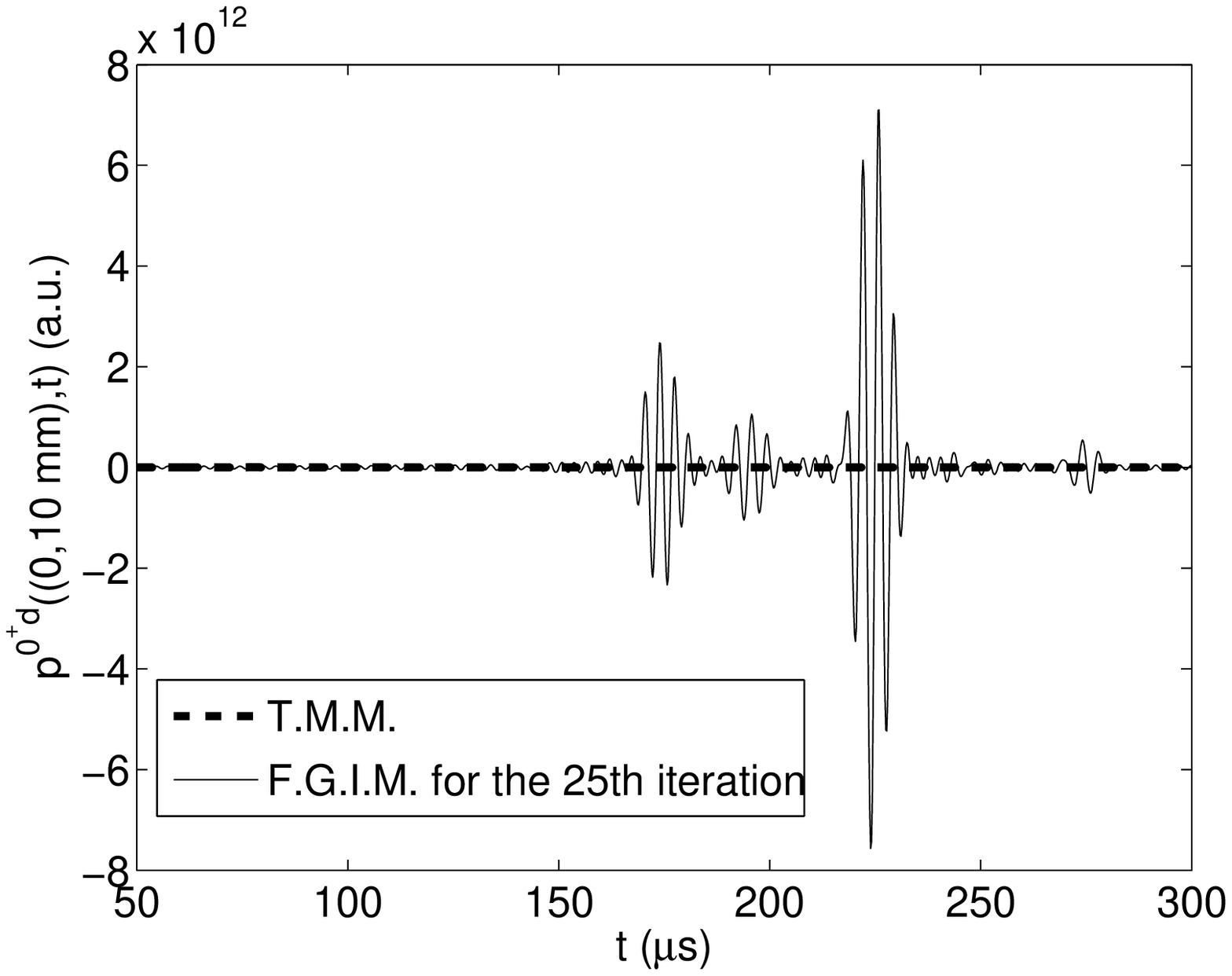,width=5.5cm}
\end{minipage}\hfill
\caption{Reflected pressure as computed by the classical Transfert Matrix Method (TMM) and as computed by the classical free-space Green's function based iterative scheme (FGIM). On the left :
the incidence angle is $0$ ; in the middle : the incidence angle is $\displaystyle \frac{\pi}{6}$ and on the right : the incidence angle is $\displaystyle \frac{\pi}{3}$.}
\label{section5f1bo1}
\end{figure}

\begin{figure}[H]
\centering\psfig{figure=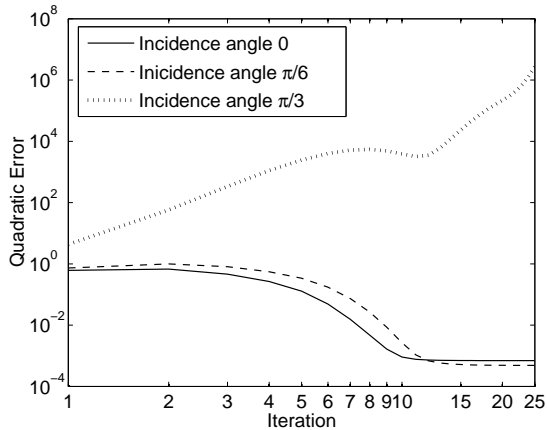,width=8.0cm}
\caption{Evolution of the quadratic errors as a function of the number
of iteration for various angles of incidence in the FGIM method.}
\label{section5f1bo2}
\end{figure}

For small angles of incidence, the usual FGIM converges,
figure \ref{section5f1bo2} and \ref{section5f1bo1}, but slower and with less accuracy than
our SGIM (figure \ref{section43f3}). For large angles of incidence
(in our example, $\displaystyle  \frac{\pi}{3}$), the usual FGIM
strongly diverges, figure \ref{section5f1bo1}, while our method still rapidly converges. This
is probably caused by the fact that the first iteration is far
from the exact solution (figure \ref{section5f1bo2}). The translation
of this divergence can be appreciated in (figure \ref{section43f4}).
All the following computations are carried out  with
characteristic parameters $\rho^{1}(\omega)$ and $k^{1}(\omega)$
filling the initially-homogeneous slab chosen as the average, over
$a \leq x_2 \leq b$,  of $\rho(x_2,\omega)$ and
$k^{1}(x_2,\omega)$ respectively.
%%
%%\begin{figure}[H]
%\begin{minipage}{8.0cm}
%%\centering\psfig{figure=figure/Profrho.eps,width=8.0cm}
%\end{minipage}\hfill
%\begin{minipage}{8.0cm}
%%\centering\psfig{figure=figure/Profk.eps,width=8.0cm}
%\end{minipage}
%%\caption{On the left: $\rho(x_2,175kHz)$. On the right: $k(x_2,175kHz)$.}
%\label{section5f1}
%%\end{figure}

\begin{figure}[H]
\begin{minipage}{0.5cm}
\centering{\rotatebox{90}{   }}
\end{minipage}\hfill
\begin{minipage}{8.0cm}
\centering{Incidence angle of $\displaystyle 0$}
\end{minipage}\hfill
\begin{minipage}{8.0cm}
\centering{Incidence angle of $\displaystyle \frac{\pi}{3}$}
\end{minipage}
\begin{minipage}{0.5cm}
\centering{\rotatebox{90}{Convergence criterion}}
\end{minipage}\hfill
\begin{minipage}{8.0cm}
\centering\psfig{figure=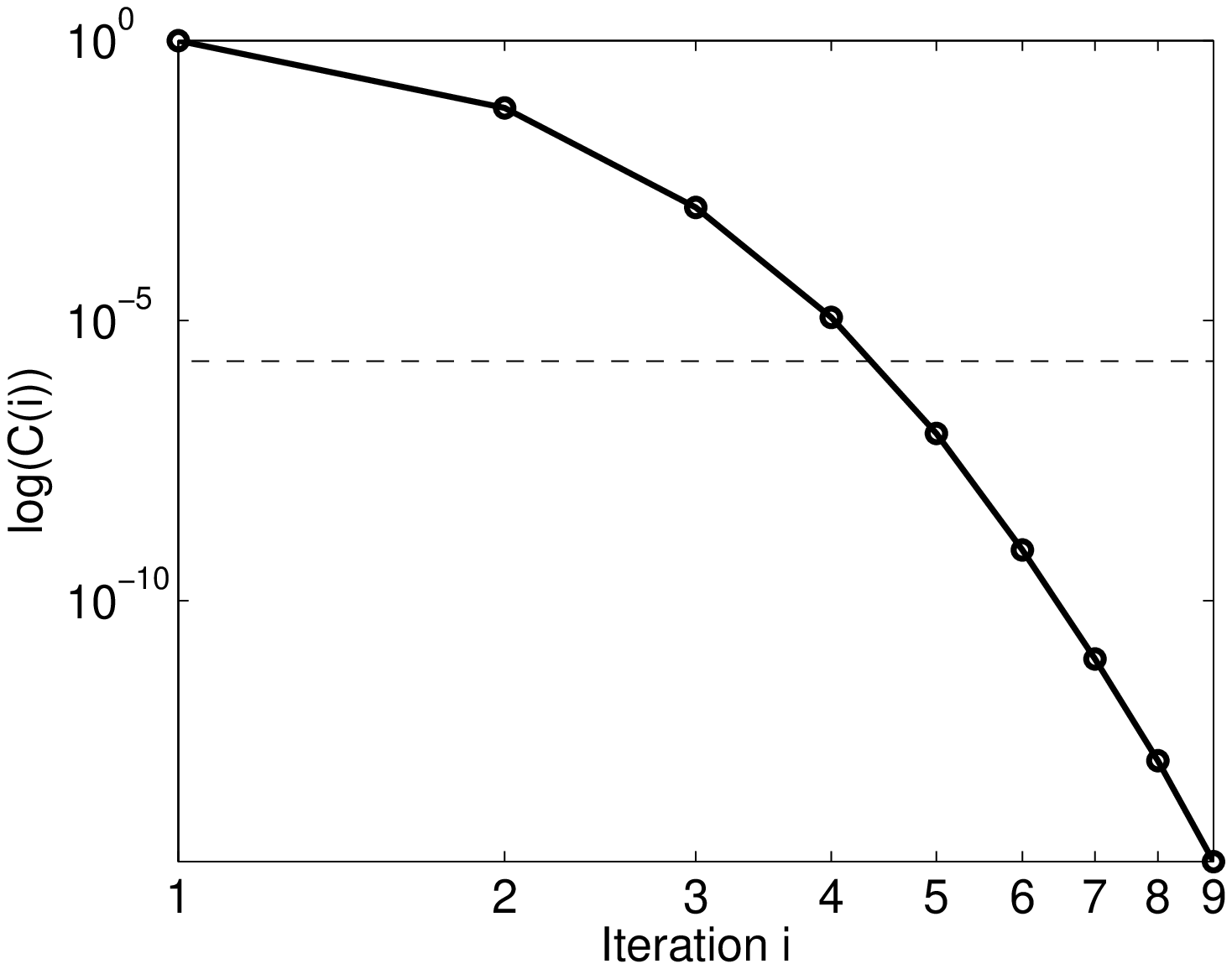,width=7.0cm}
\end{minipage}\hfill
\begin{minipage}{8.0cm}
\centering\psfig{figure=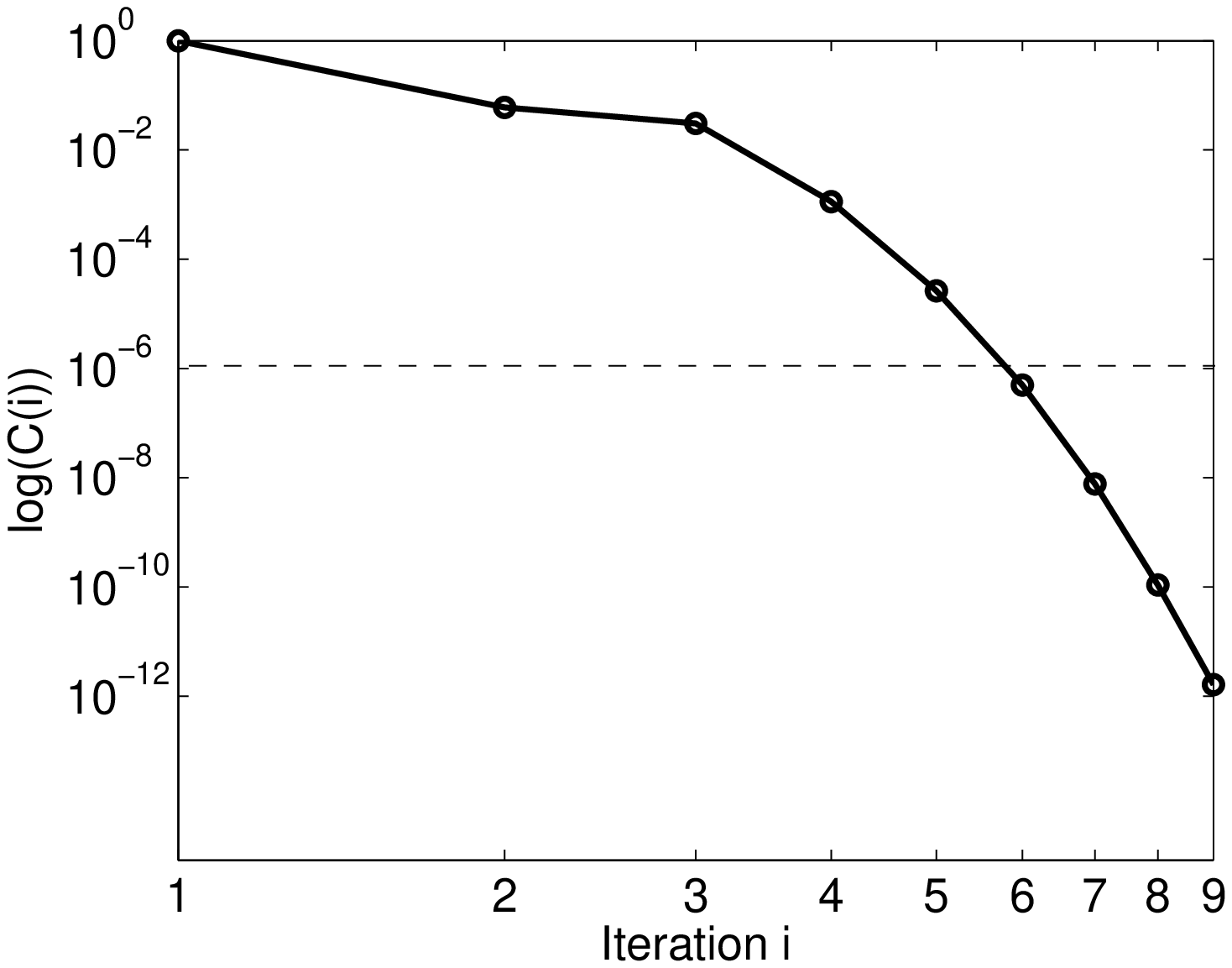,width=7.0cm}
\end{minipage}
\begin{minipage}{0.5cm}
\centering{\rotatebox{90}{Reflected pressure}}
\end{minipage}\hfill
\begin{minipage}{8.0cm}
\centering\psfig{figure=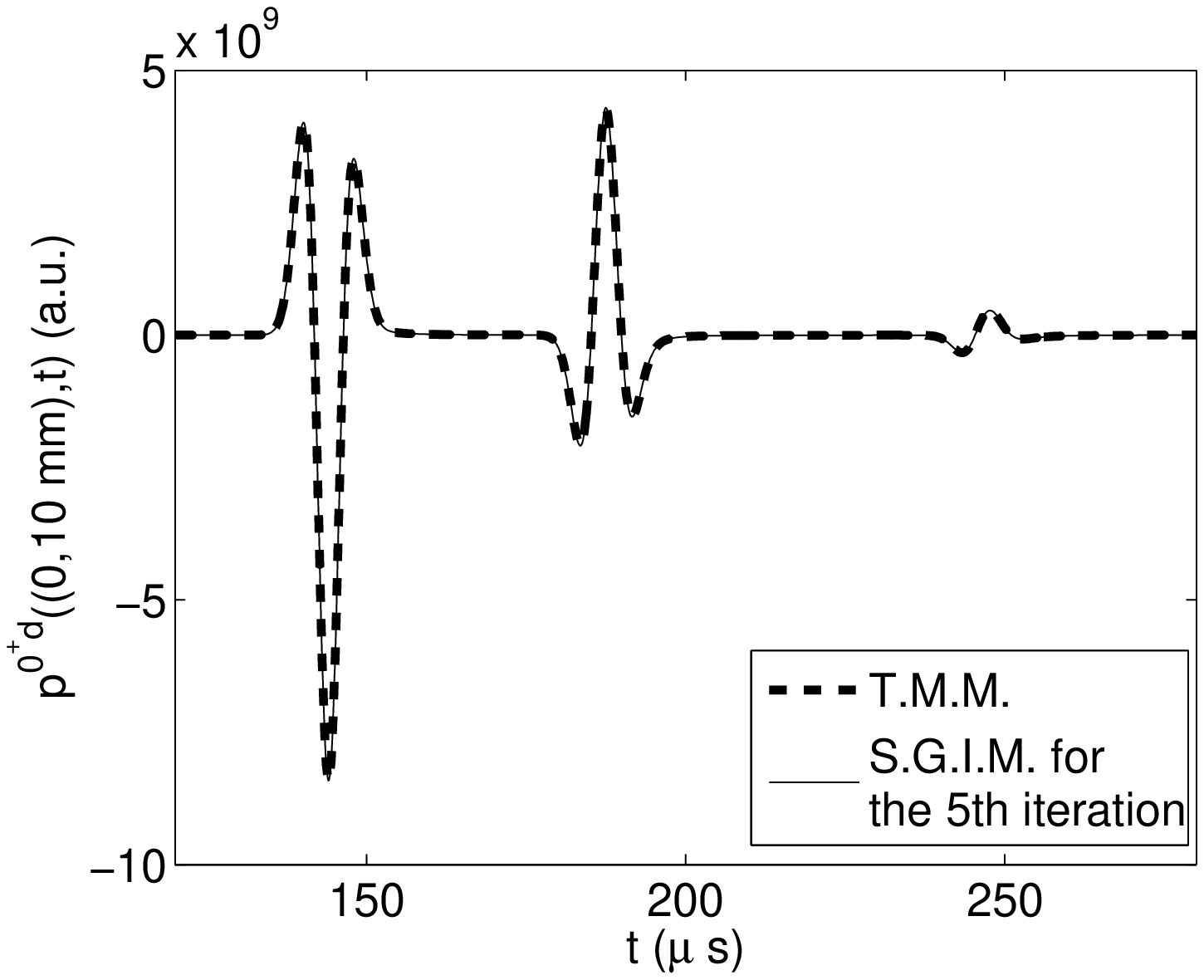,width=7.0cm}
\end{minipage}\hfill
\begin{minipage}{8.0cm}
\centering\psfig{figure=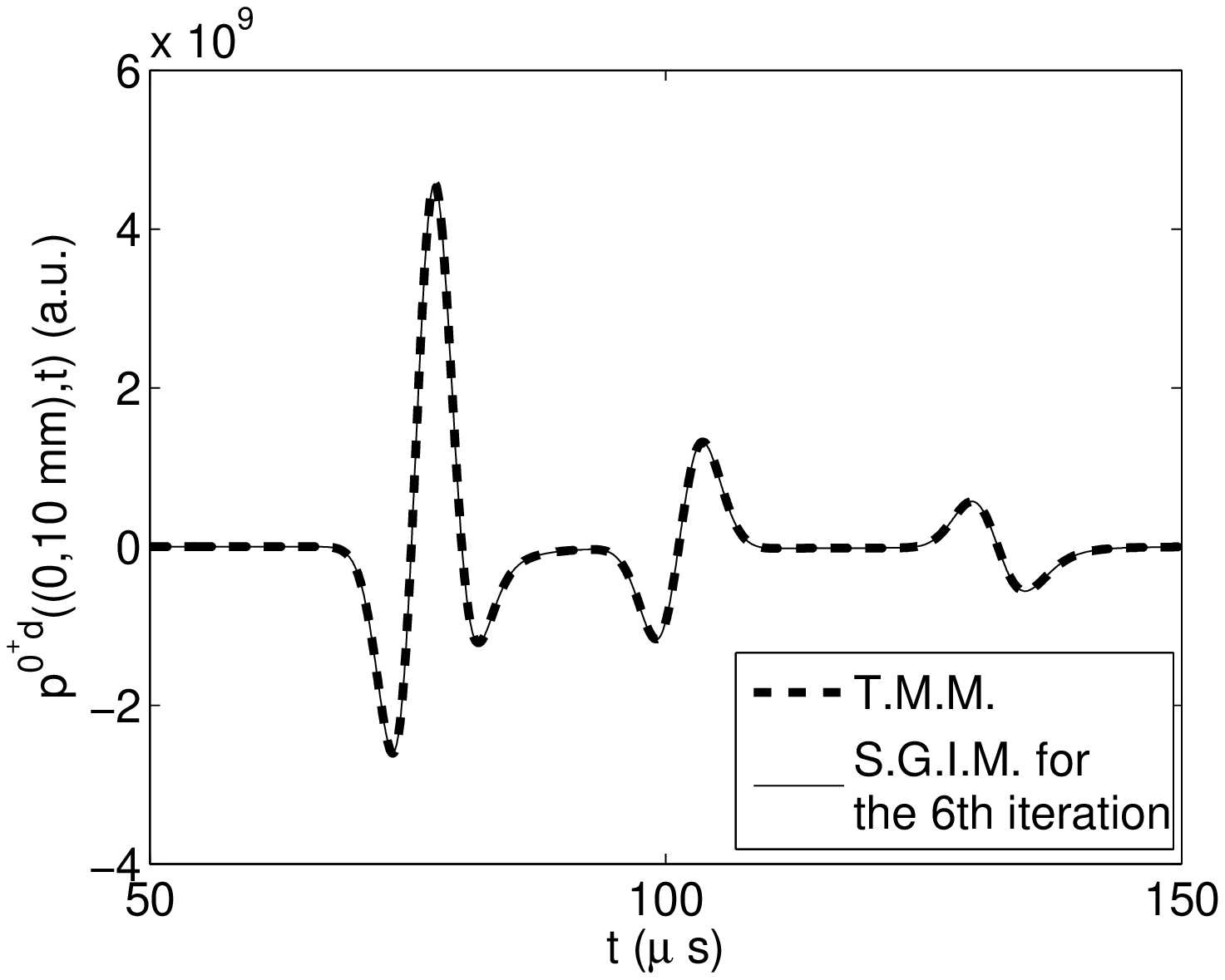,width=7.0cm}
\end{minipage}
\begin{minipage}{0.5cm}
\centering{\rotatebox{90}{Spectrum}}
\end{minipage}\hfill
\begin{minipage}{8.0cm}
\centering\psfig{figure=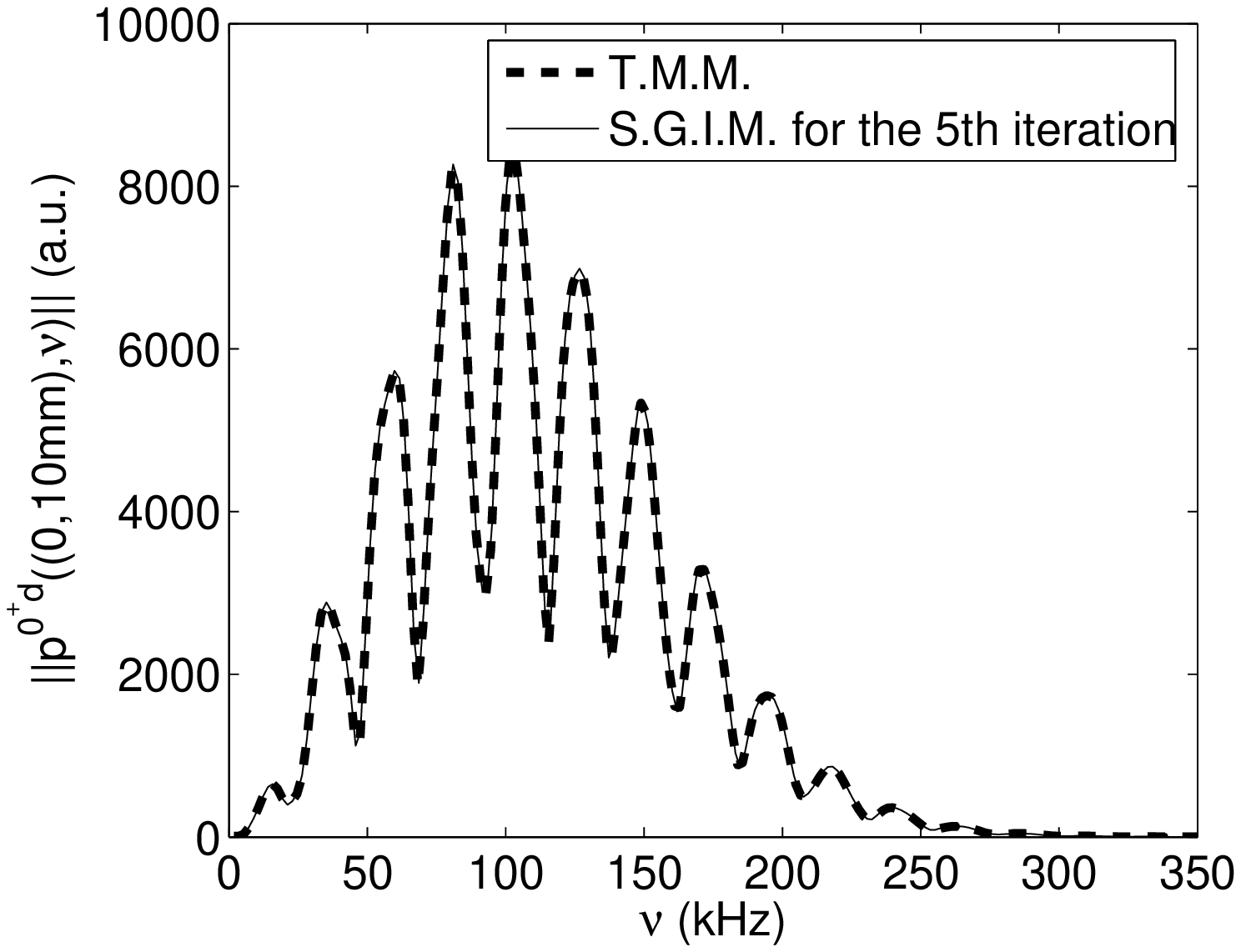,width=7.0cm}
\end{minipage}\hfill
\begin{minipage}{8.0cm}
\centering\psfig{figure=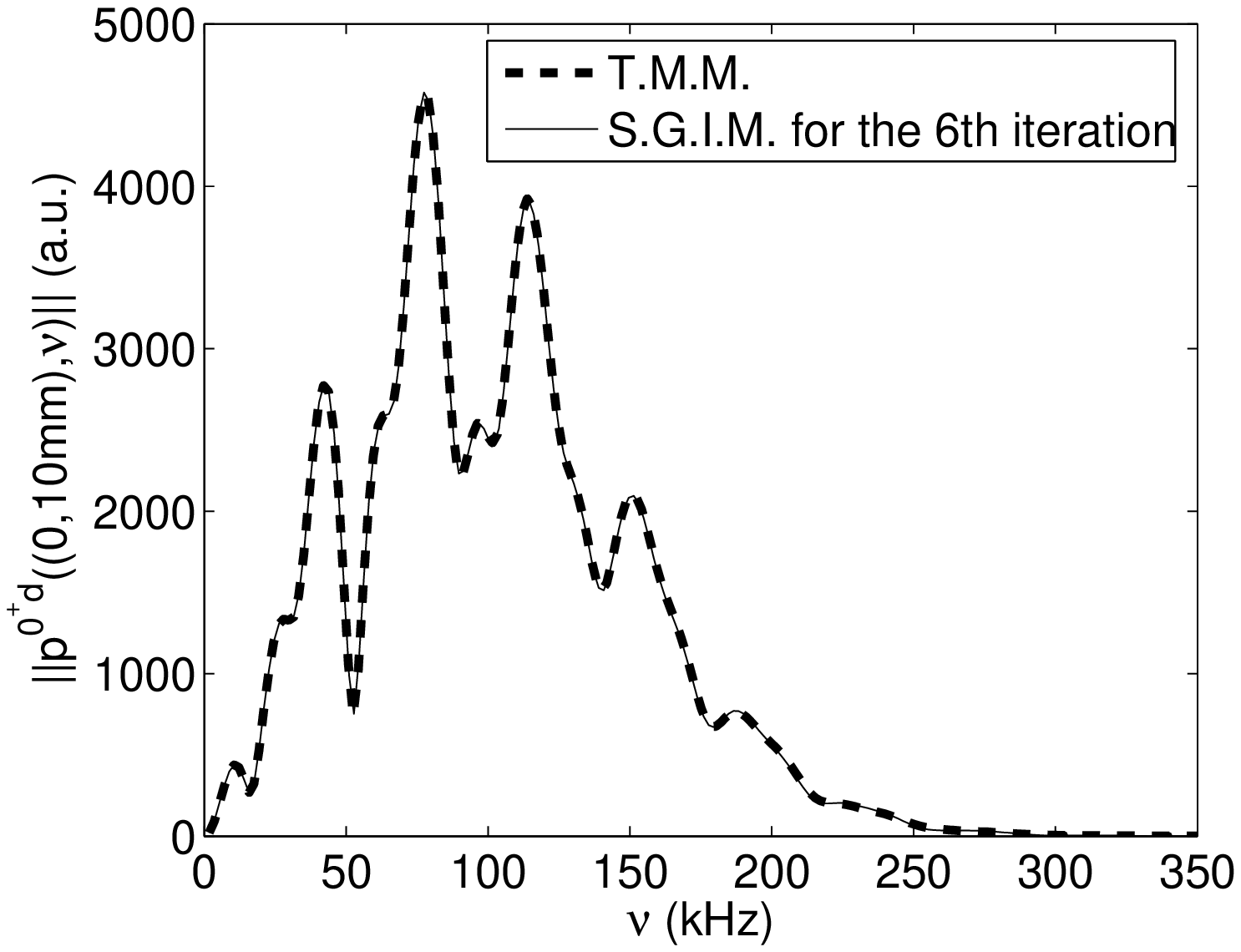,width=7.0cm}
\end{minipage}
\caption{Reflected pressure. At the top: Convergence criterion; in the middle: the reflected pressure; at the bottom: the spectrum of the reflected pressure. On the left:  incidence angle  $=0$. On the right: incidence angle $\displaystyle =\frac{\pi}{3}$.}
\label{section5f2}
\end{figure}

\begin{figure}[H]
\begin{minipage}{0.5cm}
\centering{\rotatebox{90}{   }}
\end{minipage}\hfill
\begin{minipage}{8.0cm}
\centering{Incidence angle of $\displaystyle 0$}
\end{minipage}\hfill
\begin{minipage}{8.0cm}
\centering{Incidence angle of $\displaystyle \frac{\pi}{3}$}
\end{minipage}
\begin{minipage}{0.5cm}
\centering{\rotatebox{90}{Convergence criterion}}
\end{minipage}\hfill
\begin{minipage}{8.0cm}
\centering\psfig{figure=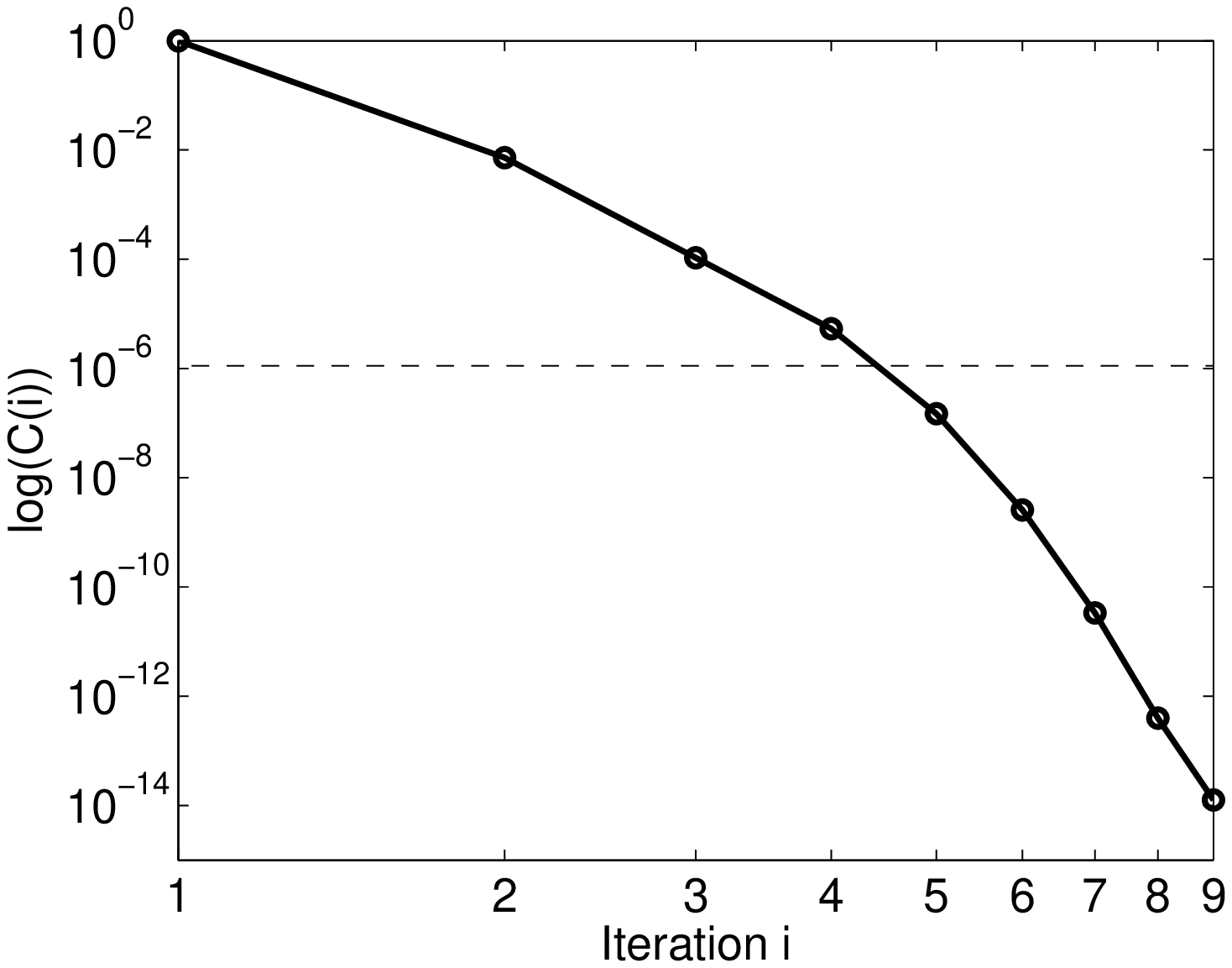,width=7.0cm}
\end{minipage}\hfill
\begin{minipage}{8.0cm}
\centering\psfig{figure=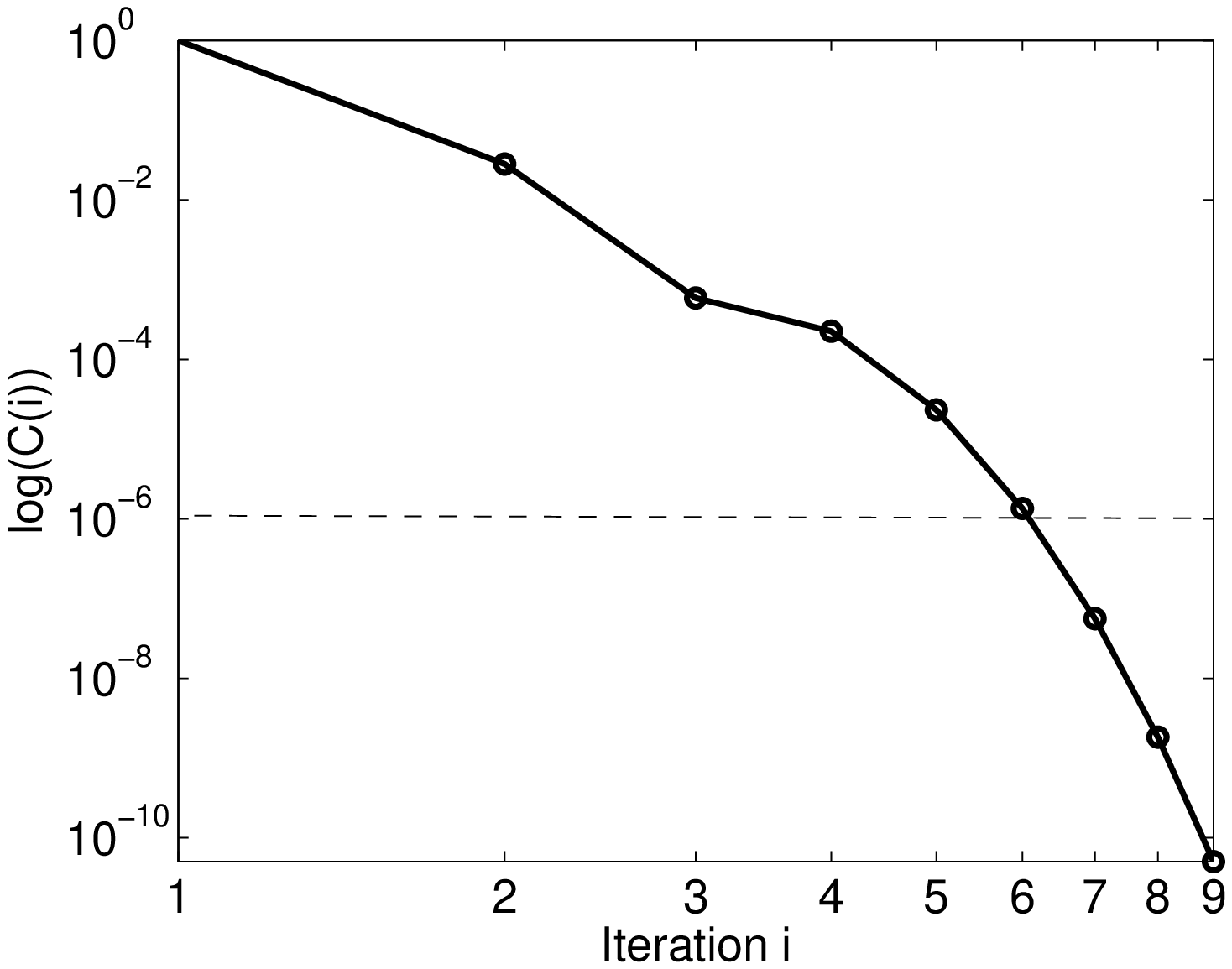,width=7.0cm}
\end{minipage}
\begin{minipage}{0.5cm}
\centering{\rotatebox{90}{Transmitted pressure}}
\end{minipage}\hfill
\begin{minipage}{8.0cm}
\centering\psfig{figure=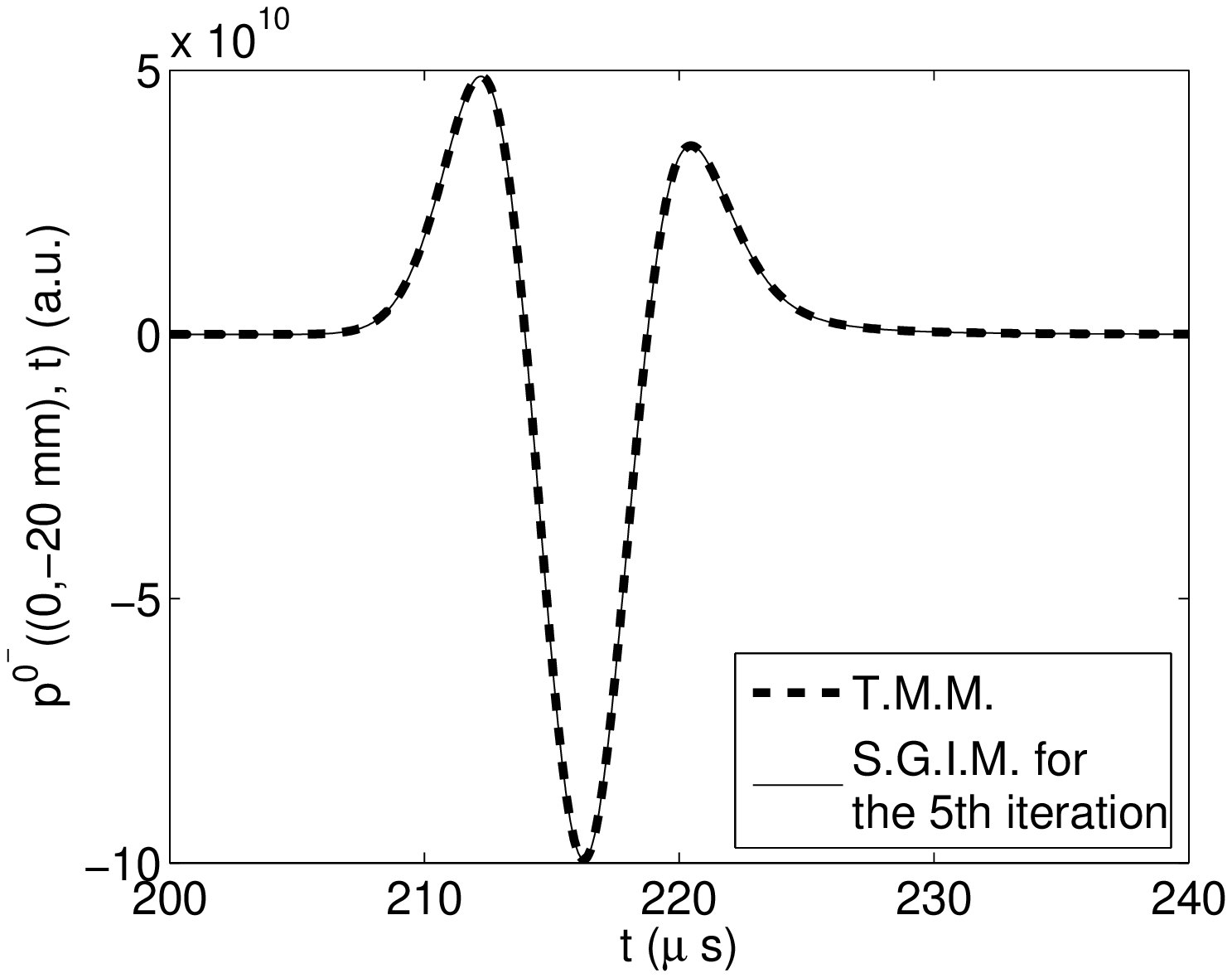,width=7.0cm}
\end{minipage}\hfill
\begin{minipage}{8.0cm}
\centering\psfig{figure=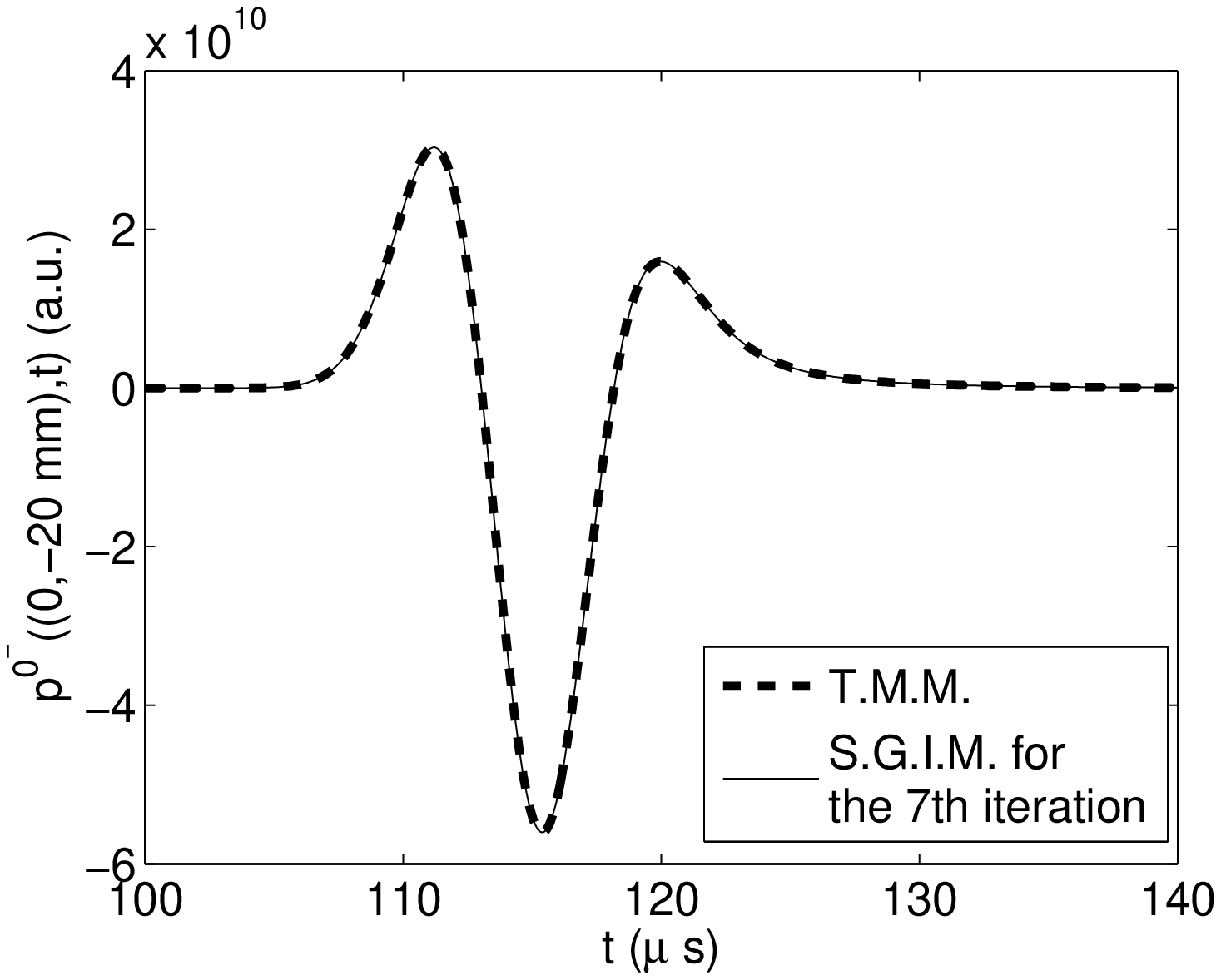,width=7.0cm}
\end{minipage}
\begin{minipage}{0.5cm}
\centering{\rotatebox{90}{Spectrum}}
\end{minipage}\hfill
\begin{minipage}{8.0cm}
\centering\psfig{figure=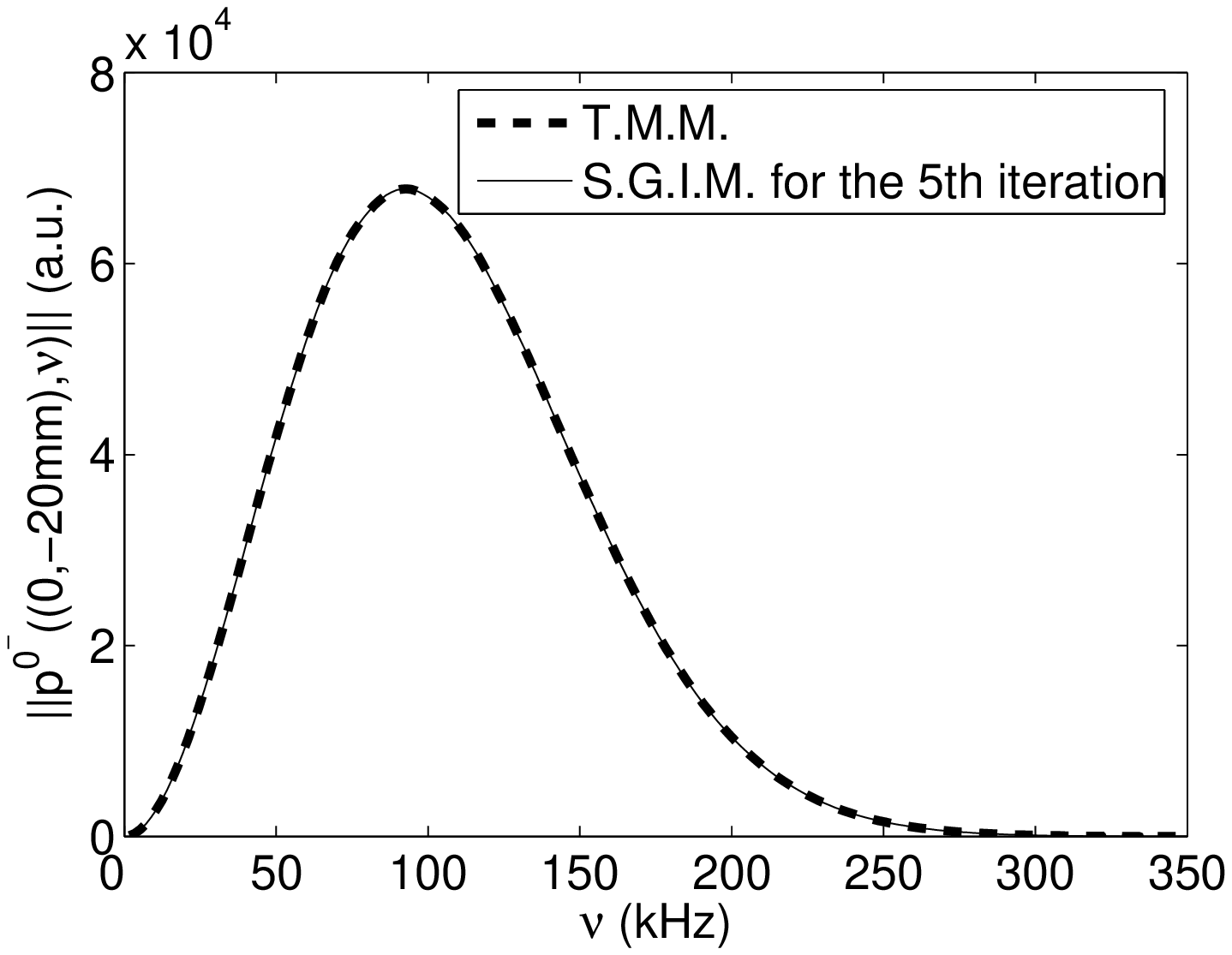,width=7.0cm}
\end{minipage}\hfill
\begin{minipage}{8.0cm}
\centering\psfig{figure=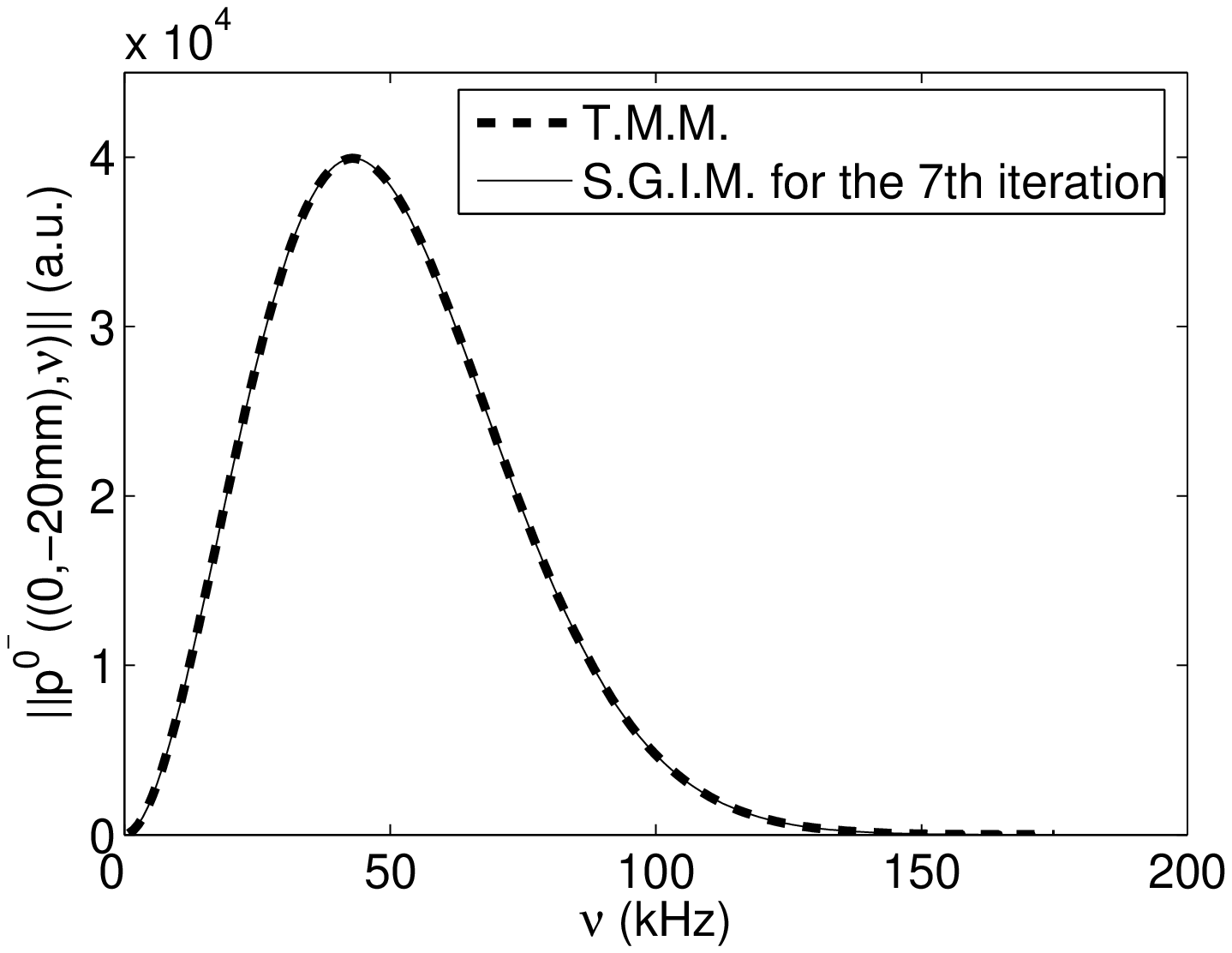,width=7.0cm}
\end{minipage}
\caption{Transmitted pressure. At the top: Convergence criterion; in the middle: the transmitted pressure; at the bottom: the spectrum of the transmitted pressure. On the left:  incidence angle  $=0$. On the right: incidence angle $\displaystyle =\frac{\pi}{3}$.}
\label{section5f3}
\end{figure}

We define a convergence criterion via the quadratic difference
between two iterations $i$ and $j$ :
\begin{equation}
D_{i,j}=\frac{\int_{0}^{T}\left(p^{0^+d,i}(\mathbf{x},t)-p^{0^+d,j}(\mathbf{x},t)
\right)^2 dt} {\int_{0}^{T}\left(p^{0^+d,j}(\mathbf{x},t)\right)^2
dt}~.
\label{section5e1}
\end{equation}
We  found empirically that a convergence criterion
$\mathcal{C}(i)$ of the form
\begin{equation}
\mathcal{C}(i) \mbox{ is true when }\frac{D_{i+1,i}}{D_{2,1}}\leq 1\times 10^{-6}
\label{section5e2}
\end{equation}
gives good results as shown figure \ref{section5f2} and
\ref{section5f3}. Our method yields solutions that are close to
those of the usual TMM both in transmission and reflection in both
the frequency and the time domain for several angles of incidence.
This validates the method employing the SGF, combined with an
iterative scheme initialized by a zeroth-order Born approximation.

The time history of the reflected pressure figure \ref{section5f2} is
of particular interest for the demonstration of the accuracy of
our method. We can clearly distinguish, for both angles of
incidence $0$ and $\displaystyle \frac{\pi}{3}$, the three
reflections of the incident wave on the three interfaces of our
canonical configuration. The zeroth-order Born approximation, i.e.
corresponding to the homogeneous fluid-saturated porous slab,
formally accounts for two of them. For the first and third
reflection, the amplitude matches correctly with our convergence
criterion. The second reflection, which comes from the
inhomogeneous slab, matches in both amplitude and time of arrival.

The time history of the transmitted pressure figure \ref{section5f3}
contains only one peak due to the fact that absorption within the
slab attenuates the transmitted waves resulting from multiple
reflection.

%%%%%%%%%%%%%%%%%%%%%%%%%%%%%%%%%%%%%%%%%%%%%%%%%%%%%%%%%%%%%%%%%%%%%%%%%%%%%%
%Ajout
\section{Conclusion}
A method, making use, in the domain integral formulation, of the
specific Green's function (SGF), i.e., the Green's function of a
canonical problem close to the original problem,  for the
resolution of problems of acoustic wave propagation in an
inhomogeneous fluid medium (with spatially-varying density and
compressibility) was studied and implemented for the canonical
example of plane wave solicitation of a double layer
fluid-saturated porous slab (considered as a single inhomogeneous
slab) in the rigid frame (equivalent fluid) approximation.

A particular feature of our study is that we account for
spatially-varying density, contrary to many authors who consider
it to be constant. We also address the issue of the spatial
differentiation of the pressure field  at the boundaries of
 the inhomogeneity, which is carried by a
finite-difference scheme for  higher-than-zeroth-order Born
approximations.

Our specific Green's function iterative scheme, which is
initialized by a zeroth-order Born approximation, was shown to
converge, contrary to the iterative scheme relying on the
free-space Green's function, which is often found to be divergent.
This improvement is due to the combined effects of the use of the
SGF and to a better Born-like approximation of the field inside
the heterogeneity.

In our numerical examples, our method was found to converge within
5-7 iterations to the reference solution (obtained rigorously by
the transfer matrix method), even for an abrupt heterogeneity, and
for various choices of the acoustic parameters filling the
homogeneous slab supposed to be the initial configuration
(canonical problem) for the SGF.

The robustness of the method was also demonstrated.

Our method thus appears useful for the  resolution of inverse
problems. In such a context, some information about the geometry
and/or the mechanical properties of the objects one is looking
for, is often known. The SGF is the device by which this
information can be incorporated into the inversion procedure in a
rational manner.
%%%%%%%%%%%%%%%%%%%%%%%%%%%%%%%%%%%%%%%%%%%%%%%%%%%%%%%%%%%%%%%%%%%%%%%%%%%%%%
\appendix
\section{The wave equation in an inhomogeneous fluid medium}\label{appendixe2}
\subsection{Solution of the direct problem involving both the pressure and
its partial derivative}\label{appendixe2s1}
Wave propagation, relative to an acoustic wave in an inhomogeneous
fluid occupying a domain $\Omega=\Omega_0 \cup \Omega_1$ (the
homogeneous host medium occupies the domain $\Omega_0$ while the
inhomogeneity occupies the domain $\Omega_1$), is described by:
\begin{equation}
\begin{array}{ll}
\displaystyle \nabla \cdot \nabla p +\frac{\omega^2}{c(\mathbf{x})^2}-
\frac{\nabla \rho(\mathbf{x})}{\rho(\mathbf{x})} \nabla p=
\rho(\mathbf{x})s(\mathbf{x})&\displaystyle \mbox{; }\forall \mathbf{x} \in \Omega
\end{array}
\label{appendixe2s1e1}
\end{equation}
wherein: $\displaystyle
c(\mathbf{x})=\sqrt{\frac{1}{\kappa(\mathbf{x})\rho(\mathbf{x})}}$
is the spatially-varying velocity,  and $\kappa(\mathbf{x})$ and
$\rho(\mathbf{x})$  the spatially-varying compressibility and
density respectively of the fluid.

Applying the domain integral formulation with the  usual
free-space Green's function, leads to the domain integral
representation of the total field
\begin{equation}
p(\mathbf{y})=p^i(\mathbf{y})+\int_{\Omega_1}
G^0\left(\mathbf{y},\mathbf{x}\right) \left(\left[
\frac{\omega^2}{c(\mathbf{x})^2}-(k^0)^2 \right]p(\mathbf{y})-
\frac{\nabla \rho(\mathbf{x})}{\rho(\mathbf{x})}\cdot \nabla
p(\mathbf{x}) \right)d\Omega(\mathbf{x})~;~ \forall
\mathbf{y}\in\Omega \label{appendixe2s1e2}
\end{equation}
wherein $p^i(\mathbf{x})$ is the incident field. To obtain the
field at an arbitrary point of space, $p$ and $\nabla p$ within
$\Omega_{1}$ have to be determined. This can be done by solving
the coupled system of integral equations:
\begin{equation}
\left\{
\begin{array}{l}
\displaystyle p(\mathbf{y})=p^i(\mathbf{y})+\int_{\Omega_1}
G^0\left(\mathbf{y},\mathbf{x}\right)\left(\left[
\frac{\omega^2}{c(\mathbf{x})^2}-(k^0)^2 \right]
p(\mathbf{y})-\frac{\nabla
\rho(\mathbf{x})}{\rho(\mathbf{x})}\cdot \nabla p(\mathbf{x})
\right)d\Omega(\mathbf{x})~;~ \forall \mathbf{y}\Omega_{1}\\[10pt]
\displaystyle
\nabla_{\mathbf{y}}p(\mathbf{y})=\nabla_{\mathbf{y}}p^i(\mathbf{y})+\int_{\Omega_{1}}\nabla_{\mathbf{y}}
G^0\left(\mathbf{y},\mathbf{x}\right)\left(\left[
\frac{\omega^2}{c(\mathbf{x})^2}-(k^0)^2
\right]p(\mathbf{y})-\frac{\nabla
\rho(\mathbf{x})}{\rho(\mathbf{x})}\cdot \nabla p(\mathbf{x})
\right)d\Omega(\mathbf{x})~;~ \forall \mathbf{y}\Omega_{1}
\end{array}
\right.
\label{appendixe2s1e3}
\end{equation}
A vast literature exists on the subject of the numerical
resolution of systems of domain integral equations
\cite{harrington,kress,mikhlin,colton, atkinson,wirgin1999}.
%%%%%%%%%%%%%%%%%%%%%%%%%%%%%%%%%%%%%%%%%%%%%%%%%%%%%%%%%%%%%%%%%%%%%%%%%%
\subsection{Solving the direct problem via a single integral equation}\label{appendixe2s2}
Another, perhaps simpler (although unsuitable in the inverse
problem context) way, to solve the previous problem is to make the
substitution
\begin{equation}
p(\mathbf{x})=q(\mathbf{x})\sqrt{\rho(\mathbf{x})}
\label{appendixe2s2e1}
\end{equation}
whereby the following governing equation is obtained
\begin{equation}
\nabla^2
q(\mathbf{x})+\left[\frac{\omega^2}{c(\mathbf{x})^2}+\frac{1}{2}\frac{\nabla^2
\rho(\mathbf{x})}{\rho(\mathbf{x})}-\frac{3}{4}
\frac{\nabla\rho(\mathbf{x})\cdot \nabla\rho(\mathbf{x})}{\left(
\rho(\mathbf{x}) \right)^2} \right]q(\mathbf{x})=
\rho^{\frac{1}{2}}(\mathbf{x})s(\mathbf{x})\mbox{  ;  }
\mathbf{x}\in\Omega~. \label{appendixe2s2e2}
\end{equation}
Using the free-space Green's function in the domain integral
formulation yields the representation
\begin{multline}
q(\mathbf{y})=q^i(\mathbf{y})+
\\
\int_{\Omega_1}
G^0\left(\mathbf{y},\mathbf{x} \right)
\left[\frac{\omega^2}{\left(c(\mathbf{x}) \right)^{2}}-\left(
k^0\right)^2+\frac{1}{2}\frac{\nabla\cdot \nabla
\rho(\mathbf{x})}{\rho(\mathbf{x})}-\frac{3}{4}
\frac{\nabla\rho(\mathbf{x})\cdot\nabla\rho(\mathbf{x})}{\left(
\rho(\mathbf{x}) \right)^2} \right] q(\mathbf{x})
dv(\mathbf{x})\mbox{  ;  }\forall \mathbf{y} \in \Omega,
\label{appendixe2s2e3}
\end{multline}
from which is extracted a single integral equation for
$q(\mathbf{x})~;~\mathbf{x}\in\Omega_{1}$.
\newline
\newline
{\textbf{Remark:}} The integral formulations
(\ref{appendixe2s1e3}) and (\ref{appendixe2s2e3}) are identical
when the density is constant.
%%%%%%%%%%%%%%%%%%%%%%%%%%%%%%%%%%%%%%%%%%%%%%%%%%%%%%%%%%%%%%%%%%%%%%%%%%%%%%
\subsection{A canonical problem involving a density discontinuity.}\label{appendixe2s3}
\begin{figure}[H]
\centering\psfig{figure=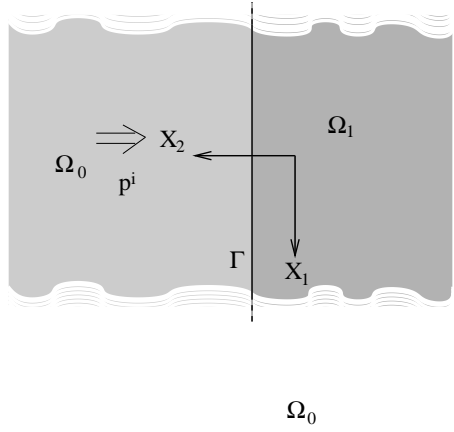,width=6.0cm}
\caption{Configuration of a planar interface between two domains.}
\label{appendixe2s3f1}
\end{figure}

Let us  consider the simple $1D$ problem, depicted in
figure \ref{appendixe2s3f1}, of a plane wave striking a planar
interface $\Gamma$, located at $x_2=a$, between two homogeneous
media $\Omega_0$ and $\Omega_1$. The normally-incident plane wave
travels initially in $\Omega_0$. The heterogeneity is supposed to
be the domain $\Omega_1$.

In practice, this problem can be treated rigorously by the TMM
method. However, when one attempts to solve it by the integral
method, the medium filling $\Omega_1$ must be dissipiative.

Let us suppose that the pressure field $p^{1}$ in $\Omega_1$ is
known (for example, calculated by the TMM method).
We introduce $\rho(x_2)=\rho^1+\left(\rho^0-\rho^1\right)H(x_2)$ and
$k(x)=k^1+\left(k^0-k^1\right)H(x_2)$, where $H(x_2)$ is the Heaviside function, and
$k^j$, $j=0,1$ the wavenumber in the domain $\Omega_j$.

Eq. (\ref{appendixe2s1e2}) splits into:
\begin{equation}
p(y_2)=p^i(y_2)+\int_{-\infty}^{a} G^0\left(y_2,x_2\right)\left(k^{1}\right)^2 p(\mathbf{y})dx_2-
\left(\rho^0-\rho^1 \right)\int_{-\infty}^{a}\frac{\delta(x_2-a)}{\rho(x_2)}
\frac{\partial p(x_2)}{\partial x_2}dx_2, \forall \mathbf{y}\Omega
\label{appendixe2s3e1}
\end{equation}
wherein $\delta$ is the Dirac delta distribution. All the
integrals involved in (\ref{appendixe2s3e1}) can be solved
analytically ($p(x_2)$ begin known by hypothesis). In particular,
the function $\displaystyle \frac{1}{\rho(x_2)}\frac{\partial
p(x_2)}{\partial x_2}$ is continuous (i.e. the function is
$C^{0}$) at the interface $\Gamma$ so that the second integral
does not present any difficulties.

The formulation involving the evaluation of $p$ and of $\nabla p$
allows us to take into account density discontinuities.  Eq.
(\ref{appendixe2s2e3}) splits into
\begin{multline}
q(y_2) =q^i(y_2)+\int_{-\infty}^{a} G^0\left(y_2,x_2\right) \left(
k^1\right)^2q(x_2) dx_2+
\\
\frac{\left(\rho^0-\rho^1 \right)}{2}
\int_{-\infty}^{a} \frac{\delta'(x_2-a)}{\rho(x_2)}q(x_2) dx_2 -
\\
\frac{3\left(\rho^0-\rho^1 \right)^2}{4}  \int_{-\infty}^{a}
\frac{ \delta(x_2-a) \delta(x_2-a)}{\left( \rho(\mathbf{x})
\right)^2} q(x_2) dx_2\mbox{  ;  }\forall y_2 \in \Omega~.
\label{appendixe2s3e2}
\end{multline}
The calculation of first integral presents no particular
difficulties. Let us consider the second integral $\displaystyle
\int_{-\infty}^{a}\frac{\delta'(x_2-a)}{\rho(x_2)}q(x_2) dx_2 $,
wherein $\delta'(x_2-a)$ is the derivative of the Dirac delta
distribution. The use of the formula
\begin{equation}
\int f(x_2) \delta'(x_2-a) dx_2=-\frac{\partial f(x_2)}{\partial x_2}\delta(x_2-a)
\label{appendixe2s3e3}
\end{equation}
requires the function $f(x_2)$ to be $C^{1}$ at $x_2=a$, while the
function $\displaystyle
\frac{q(x_2)}{\rho(x_2)}=\frac{p(x_2)}{\rho(x_2)^{\frac{3}{2}}}$
is not continuous at the interface $\Gamma$. Thus, this second
term cannot be handled analytically.

Finally,  consider the third term $\displaystyle
\int_{-\infty}^{a} \frac{ \delta(x_2-a)\delta(x_2-a)}{\left(
\rho(\mathbf{x}) \right)^2} q(x_2) dx_2$. The integrand involves
the scalar product of two Dirac delta distributions
$\delta(x_2-a) \delta(x_2-a)$ which is not defined
\cite{Kurasov,Colombeau1992,Colombeau1984}. A Numerical
approximation of this quantity exists, but otherwise it is
meaningless \cite{Colombeau1984}.

The resolution of problems via a single equation is also of no
practical use when the problem one is faced with involves density
discontinuities.
%Afin ajout
%%%%%%%%%%%%%%%%%%%%%%%%%%%%%%%%%%%%%%%%%%%%%%%%%%%%%%%%%%%%%%%%%%%%%%%%%%%%%%
\section{Pressure field in the case of a macroscopically-homogeneous porous slab (zeroth-order Born approximation).}
\label{appendixe1}
By referring to \cite{groby2005}, one finds, that for  plane wave
solicitation in $\Omega_0^+$, the pressure fields in $\Omega_0^-$,
$\Omega_1$ and $\Omega_0^+$ are:
\begin{multline}
p^{0+}(\mathbf{x},\omega)=A^i(\omega)\exp\left(\mbox{i}k_1^i
x_1-\mbox{i}k_2^{0,i}x_2 \right)+
\\
A^i(\omega)\frac{\mbox{i} \exp\left( \mbox{i}k_1^i
x_1+\mbox{i}k_2^{0,i}(x_2-2a)\right)\sin\left(k_2^{1,i} l
\right)\left((\alpha^{1,i})^2-(\alpha^{0,i})^2 \right)}{2\alpha^{1,i} \alpha^{0,i}
\cos\left(k_2^{1,i} l
\right)-\mbox{i}\left((\alpha^{1,i})^2+(\alpha^{0,i})^2
\right)\sin\left(k_2^{1,i} l \right)}=\exp\left(\mbox{i}k_1^i x_1
\right)\widetilde{p}^{0-}(x_2,\omega)
\label{appendix1e1}
\end{multline}
\begin{multline}
p^{1}(\mathbf{x},\omega)=
A^i(\omega)\frac{2\exp\left(\mbox{i}k_1^i x_1-\mbox{i}k_2^{0,i}a
\right)\alpha^{0,i}\left[\alpha^{1,i}\cos\left(k_2^{1,i} (x_2-b)
\right)-\mbox{i}\alpha^{0,i}\sin\left(k_2^{1,i} (x_2-b)
\right)\right]}{2\alpha^{1,i} \alpha^{0,i} \cos\left(k_2^{1,i} l
\right)-\mbox{i}\left((\alpha^{1,i})^2+(\alpha^{0,i})^2
\right)\sin\left(k_2^{1,i} l \right)}=
\\
\exp\left(\mbox{i}k_1^i x_1 \right)\widetilde{p}^{1}(x_2,\omega)
\label{appendix1e2}
\end{multline}
\begin{multline}
\frac{\partial p^{1}(\mathbf{x},\omega)}{\partial x_2}=
A^i(\omega)\frac{2\exp\left(\mbox{i}k_1^i x_1-\mbox{i}k_2^{0,i}a
\right)\alpha^{0,i}k_{2}^{1,i}\left[-\alpha^{1,i}\sin\left(k_2^{1,i} (x_2-b)
\right)-\mbox{i}\alpha^{0,i}\cos\left(k_2^{1,i} (x_2-b)
\right)\right]}{2\alpha^{1,i} \alpha^{0,i} \cos\left(k_2^{1,i} l
\right)-\mbox{i}\left((\alpha^{1,i})^2+(\alpha^{0,i})^2
\right)\sin\left(k_2^{1,i} l \right)}=
\\
\exp\left(\mbox{i}k_1^i x_1 \right)\frac{\partial\widetilde{p}^{1}(x_2,\omega)}{\partial x_2}
\label{appendix1e2b}
\end{multline}
\begin{equation}
p^{0-}(\mathbf{x},\omega)=A^i(\omega)\frac{2\exp\left(\mbox{i}k_1^i
x_1-\mbox{i}k_2^{0,i}\left( x_2+l\right) \right)\alpha^{1,i} \alpha^{0,i}
}{2\alpha^{1,i} \alpha^{0,i} \cos\left(k_2^{1,i} l
\right)-\mbox{i}\left((\alpha^{1,i})^2+(\alpha^{0,i})^2
\right)\sin\left(k_2^{1,i} l \right)}=\exp\left(\mbox{i}k_1^i x_1
\right)\widetilde{p}^{0-}(x_2,\omega)~.
\label{appendix1e3}
\end{equation}
%%
%%%%%%%%%%%%%%%%%%%%%%%%%%%%%%%%%%%%%%%%%%%%%%%%%%%%%%%%%%%%%%%%%%%%%%%%%%%%%%
\section{Pressure field in the case of a double layer macroscopically-homogeneous porous slabs.}
\label{appendixe3}
We use a separation of variables technique to obtain the field representations:
\begin{equation}
\begin{array}{l}
\displaystyle p^{0^+}=A^{i}(\omega)e^{\mbox{i}[k_1^i x_1-k_2^{0,i}x_2]}+B^{0^+}e^{\mbox{i}[k_1^i x_1+k_2^{0,i}\left(x_2-a\right)]}\\[8pt]
\displaystyle  p^{1}=e^{\mbox{i}k_1^i x_1}\left(A^{1}e^{-\mbox{i}k_2^{1,i}\left(x_2-a\right)}+B^{1}e^{\mbox{i}k_2^{1,i}\left(x_2-a\right)}\right)\\[8pt]
\displaystyle p^{2}=e^{\mbox{i}k_1^i x_1}\left(A^{2}e^{-\mbox{i}k_2^{2,i}\left(x_2-b\right)}+B^{2}e^{\mbox{i}k_2^{2,i}\left(x_2-b\right)}\right)\\[8pt]
\displaystyle p^{0^{-}}=A^{0^-}e^{\mbox{i}[k_1^i x_1-k_2^{0,i}\left(x_2-b\right)]}
\end{array}
\label{appendixe3e1}
\end{equation}
After introducing the fields expressions into the boundary conditions (continuity of the pressure and of the normal velocity), we multiply these relations by $\exp\left(-\mathbf{i}K_1x_1 \right)$ and then integrate form $-\infty$ to $+\infty$ to obtain the matrix equation (solved numerically to get $B^{0^+}$ and $A^{0^-}$)
\begin{equation}
\left(\!
\begin{array}{llllll}
\displaystyle 1&\displaystyle -1&\displaystyle -1&\displaystyle 0&\displaystyle 0&\displaystyle 0\\[8pt]
\displaystyle \alpha^{0,i}&\displaystyle \alpha^{1,i}&\displaystyle -\alpha^{1,i}&\displaystyle 0&\displaystyle 0&\displaystyle 0\\[8pt]
\displaystyle 0&\displaystyle e^{\mbox{i}k_2^{1,i}l^1}&\displaystyle e^{-\mbox{i}k_2^{1,i}l^1}&\displaystyle -e^{-\mbox{i}k_2^{2,i}l^2}&\displaystyle -e^{\mbox{i}k_2^{2,i}l^2}&\displaystyle 0\\[8pt]
\displaystyle 0&\displaystyle -\alpha^{1,i} e^{\mbox{i}k_2^{1,i}l^1}&\displaystyle \alpha^{1,i} e^{-\mbox{i}k_2^{1,i}l^1}&\displaystyle \alpha^{2,i} e^{-\mbox{i}k_2^{2,i}l^2}&\displaystyle -\alpha^{2,i} e^{\mbox{i}k_2^{2,i}l^2}&\displaystyle 0\\[8pt]
\displaystyle 0&\displaystyle 0&\displaystyle 0&\displaystyle 1&\displaystyle 1&\displaystyle -1\\[8pt]
\displaystyle 0&\displaystyle 0&\displaystyle 0&\displaystyle -\alpha^{2,i}&\displaystyle\alpha^{2,i}&\displaystyle \alpha^{0,i}\\[8pt]
\end{array}\!
 \right)
\!
\left(\!
\begin{array}{l}
\displaystyle  \!B^{0^+} \!\\[8pt]
\displaystyle  \!A^{1}\!\\[8pt]
\displaystyle  \!B^{1}\!\\[8pt]
\displaystyle  \!A^{2}\!\\[8pt]
\displaystyle  \!B^{2}\!\\[8pt]
\displaystyle  \!A^{0^-}\!\\[8pt]
\end{array}\!
\right)
\!
=\!
\left(\!
\begin{array}{l}
\displaystyle \!-A^{i}(\omega)e^{-\mbox{i}k_2^{0,i}a} \!\\[8pt]
\displaystyle  \!\alpha^{0,i}A^{i}(\omega)e^{-\mbox{i}k_2^{0,i}a} \!\\[8pt]
\displaystyle 0\\[8pt]
\displaystyle 0\\[8pt]
\displaystyle 0\\[8pt]
\displaystyle 0\\[8pt]
\end{array}\!
\right)
\label{appendixe3e2}
\end{equation}
wherein $l^1$ and $l^2$ are the thickness of the layer 1 and the layer 2  respectively (table \ref{tab1}) and  $\displaystyle \alpha^{j,i}=\frac{k_2^{j,i}}{\rho^j}$, $j=0^+,1,2,0^-$.
%---------------------------------------------------------------
\bibliography{biblio}
\bibliographystyle{plain}
\end{document}